\documentclass[letterpaper,11pt]{article}
\pdfoutput=1 % to force arxiv to pdflatex

\usepackage{jheppub}
\usepackage{graphicx}
\usepackage{hyperref}
\usepackage{caption,subcaption} % Subfigures
\usepackage[section]{placeins}
\usepackage[utf8]{inputenc}

\newcommand{\be}{\begin{equation}}
\newcommand{\ee}{\end{equation}}
\newcommand{\bea}{\begin{eqnarray}}
\newcommand{\eea}{\end{eqnarray}}
\newcommand{\R}{\mathcal{R}}
\newcommand{\mass}{\mathcal{M}}
\newcommand{\comment}[1]{}

\title{Black Hole Formation in AdS Einstein-Gauss-Bonnet Gravity}

\author[a]{Nils Deppe}
\author[b]{Allison Kolly}
\author[c]{Andrew R.~Frey}
\author[c]{and Gabor Kunstatter}

\affiliation[a]{Cornell Center for Astrophysics and Planetary Science
and Department of Physics\\ Cornell University\\
122 Sciences Drive, Ithaca, New York 14853, USA}

\affiliation[b]{Department of Atmospheric and Oceanic Sciences, McGill
University\\ 805 Sherbrooke Street West\\
Montr\'eal, Qu\'ebec H3A 0B9, Canada}

\affiliation[c]{Department of Physics and Winnipeg Institute for Theoretical
Physics, University of Winnipeg\\
515 Portage Avenue, Winnipeg, Manitoba R3B 2E9, Canada}

\emailAdd{nd357@cornell.edu}
\emailAdd{allison.kolly@mail.mcgill.ca}
\emailAdd{a.frey@uwinnipeg.ca}
\emailAdd{g.kunstatter@uwinnipeg.ca}

\abstract{AdS spacetime has been shown numerically to be unstable against a 
large class of arbitrarily small perturbations.  In \cite{Deppe:2014oua}, 
the authors 
presented a preliminary study of the effects on stability of changing the 
local dynamics by adding a Gauss-Bonnet term to the Einstein action. 
Here we provide further details as well as new results with improved numerical
methods. In particular, we elucidate new structure in Choptuik scaling plots. 
We also provide evidence of chaotic behavior at the transition between 
immediate horizon formation and horizon formation after the matter
pulse reflects from the AdS conformal boundary.
Finally, we present data suggesting 
the formation of naked singularities in spacetimes with ADM mass below the 
algebraic bound for black hole formation.}


\begin{document}
\maketitle

\section{Introduction}\label{s:intro}

It is well known that, in Einstein gravity, Minkowski space is stable against 
arbitrarily small perturbations. The simplest way to understand this is by 
noting that the formation of microscopic black holes exhibits critical 
behaviour, usually referred to as Choptuik scaling \cite{Choptuik:1992jv}. 
Specifically, as the amplitude of a small initial perturbation gradually 
decrease, the dynamics undergoes a phase transition between black hole 
formation (for large amplitudes) and dispersion to infinity (for small 
amplitudes). Infinitesmally small perturbations invariably disperse.  As with 
all critical phenomena,  the end state is highly sensitive to small variations 
in the initial data near the transition region.
The properties of the phase transition depend both qualitatively and 
quantitatively on the nature of the local dynamics. The transition is second 
order in the absence of a fundamental scale in the problem, but can be either 
first or second order when new scales are introduced. New scales can arise 
either in the matter action or due to modifications to the gravitational 
dynamics via, for example, the addition of higher curvature terms. It has been 
shown \cite{Deppe:2012wk} that adding a Gauss-Bonnet (GB) curvature squared
term in 
five and six spacetime dimensions radically affects the critical behaviour in 
microscopic black hole formation.\footnote{Choptuik scaling in 5D
Einstein-Gauss-Bonnet gravity was 
first considered in \cite{Golod:2012yt}.}

Naively one might expect Choptuik scaling, which was originally discovered 
as a local phenomenon, to be insensitive to boundary conditions at infinity. 
In particular, it was thought to be unaffected by the inclusion of a 
cosmological constant. To the contrary, \cite{Bizon:2011gg} argued from 
numerical
results that, in (global) anti-deSitter space (AdS), which exhibits reflecting 
boundary conditions at the conformal boundary, black holes form from 
arbitrarily small perturbations for massless scalar matter with
a large class of initial data.
The instability arises because subcritical matter that initially disperses 
is able to return from the boundary in finite time to form a horizon near the 
origin after additional gravitational focusing. Subsequent analysis by many
authors \cite{Garfinkle:2011hm,Jalmuzna:2011qw,Garfinkle:2011tc,Dias:2011ss,
Buchel:2012uh,Dias:2012tq,Buchel:2013uba,Maliborski:2013jca,Maliborski:2013ula,
Balasubramanian:2014cja,Bizon:2014bya,Maliborski:2014rma,Buchel:2014xwa,
Dimitrakopoulos:2014ada,
Craps2014,Craps2015,Balasubramanian:2015uua,Bizon:2015pfa,Fodor:2015eia,
Green:2015dsa,
Dimitrakopoulos:2015pwa,Craps:2015iia,Craps:2015jma,Craps:2015xya,Dias:2016ewl,
Dimitrakopoulos:2016tss,Deppe:2016gur}
has demonstrated the existence of ``islands of stability,'' i.e. 
non-negligible regions of the initial data parameter space for which black 
holes never form.  In fact, some perturbative analysis suggests that 
stability against horizon formation may be generic in the parameter space
of initial conditions, and it is still an open question whether
stability, instability, or both are technically generic at arbitrarily small
but finite amplitude.  Other work has considered massive scalars 
\cite{Buchel:2014dba,Okawa2015,Deppe:2015qsa}, a gauge field and charged
scalar \cite{Arias:2016aig}, and holographic models of confining theories
(related to the Poincar\'e patch rather than global AdS)
\cite{Craps:2014eba,Craps:2015upq}.
   
The stability of AdS spacetime is an interesting question in 
mathematical physics in its own right, but the issue takes on particular 
significance in the context of the AdS/CFT correspondence, in which  gravity 
in AdS spacetime is dual to a Yang-Mills theory on the conformal boundary.
Since black hole formation in the bulk AdS spacetime corresponds to 
thermalization in the spatially compact boundary CFT, it is perhaps less 
surprising to think that generic initial conditions lead to black holes.
Indeed, islands of stability are more surprising as they imply that 
some low-energy perturbations of Yang-Mills theories on $S^3$ need not
thermalize.  However, the high degree of symmetry in AdS (integrability of
the boundary theory) can lead to quasiperiodic behavior.  It is clear that
AdS/CFT is a rich system with many lessons about nonperturbative dynamical
behavior.
\comment{\footnote{One caveat is that the small black holes formed at small
amplitudes in AdS suffer a rapid Gregory-Laflamme-like instability over
the extra dimensions in explicit examples of the AdS/CFT correspondence.
Nonetheless, it is not yet clear that the instability sets in substantially
before formation of the small horizon} }

The end state of gravitational collapse in AdS spacetime results from the 
interplay of local (weak turbulence) and global (resonance) dynamics of the 
spacetime. Quantum theory generically suggests the need for higher-derivative 
terms in the gravitational and matter actions that necessarily alter the 
short distance, high curvature dynamics near the final stages of gravitational 
collapse, i.e. the local dynamics. In the AdS/CFT correspondence, 
higher-curvature terms in the gravitational action correspond to finite $N$
and 't Hooft coupling effects in the dual theory, including differing 
$a$ and $c$ central charges in 4D CFTs.  Our focus in this paper is the
gravitational collapse of a massless scalar field in Einstein-Gauss-Bonnet
(EGB) gravity in the AdS$_5$/CFT$_4$ correspondence.  We are motivated in
part by the possible relation of the boundary CFT to strong dynamics in
QCD.

While one expects a tower of higher-derivative couplings suppressed by powers
of the string scale, 5D EGB gravity has been an important model of 
higher-curvature effects in the AdS/CFT correspondence because it is the first
example of Lovelock gravity \cite{Lanczos1938,Lovelock1971} beyond the 
Einstein-Hilbert action.\footnote{There are no non-trivial Lovelock terms in 
4D. The GB term is a total divergence.}  The key feature of Lovelock terms in 
the gravitational action 
is that the equations of motion remain second order in derivatives of the 
metric despite the fact that the action is higher order. Not only does this 
imply that the theory is ghost-free when linearized around a flat background, 
but it also makes the study of AdS stability tractable. 
In 5D, only the lowest order Lovelock term (beyond Einstein), the Gauss-Bonnet 
term, is relevant.  As a result, it is the unique higher-curvature theory 
of gravity with second-order equations of motion.

The present authors initiated a study of the stability of AdS in EGB gravity 
in \cite{Deppe:2014oua}. The purpose of the current paper is to provide further 
details of our calculations as well as new results with improved numerical
methods. 
In particular, we present an additional discussion of structure in critical 
behavior near transitions between collapse before and after reflection from
the conformal boundary, evidence for self-similar (that is, chaotic) 
behaviour in the black hole formation time vs amplitude plots in transition
regions, and data hinting at the formation of naked singularities in 
spacetimes with ADM mass below the algebraic bound for black hole formation.

The paper is organized as follows. In section \ref{s:egb gravity},
we review EGB gravity and 
derive via Hamiltonian techniques the relevant equations of motion in 
Schwarzschild coordinates.  We also briefly describe our numerical methods 
there.  We discuss our results on the above topics in section
\ref{s:results}. 
We close with a summary and prospects for future work. An appendix contains 
the derivation of the equations of motion for the same system but in the 
AdS analogue of flat slice coordinates for future reference.

\section{EGB Gravity and EoMs}\label{s:egb gravity}

In this section, we briefly review features of Einstein-Gauss-Bonnet gravity in 
AdS$_5$ and the Hamiltonian derivation of both the mass function and scalar
equations of motion.

\subsection{Einstein-Gauss-Bonnet Gravity In AdS}

$5$-dimensional Einstein-Gauss-Bonnet gravity is a special case of 
Lanczos-Lovelock gravity \cite{Lanczos1938,Lovelock1971}. The action is
\be
I_{EGB}=\frac{1}{2\kappa_5{}^2}\int d^5x \sqrt{-g}\left({12}{\lambda}
+ {\mathcal{R}} + \frac{\lambda_3}{2}   \left[\mathcal{R}^2 - 
4 \mathcal{R}_{\mu\nu}\mathcal{R}^{\mu\nu} +
\mathcal{R}_{\mu\nu\rho\sigma}\mathcal{R}^{\mu\nu\rho\sigma}\right]
\right) \label{eq:action1}
\ee
with $\lambda>0$ in AdS (we use $\R$ for the Riemann tensor and its
contractions).
The covariant equations of motion are \cite{Nozawa2008,Maeda2011}:
\be
G_{\mu\nu} +\lambda_3 H_{\mu\nu}-\frac{\lambda}{24}g_{\mu\nu}=0\ ,
\label{eq:covariant}
\ee
where
\bea
G_{\mu\nu} &=&\R_{\mu\nu}-\frac{1}{2}g_{\mu\nu}\R  \textnormal{ and}\\
H_{\mu\nu} &=& 2\left[\R \R_{\mu\nu} - 2\R_{\mu\alpha}\R^\alpha{}_\nu- 2
\R^{\alpha\beta}\R_{\mu\alpha\nu\beta}+ \R_\mu{}^{\alpha\beta\gamma}
\R_{\nu\alpha\beta\gamma}\right]-\frac{1}{2}g_{\mu\nu}L_{GB}\ .
\label{eq:H}
\eea
A key feature of spherically symmetric EGB is the existence of a generalized 
mass function 
\be
\mass= \frac{3}{2\kappa_5^2}R^{4}\left[\lambda + \frac{1}{R^2}\left(1-
|DR|^2\right) +\frac{\lambda_3}{R^4}\left(1-|DR|^2\right)^2\right]\ ,
\label{eq:M}
\ee
where $R$ is the areal radius and
$|DR|^2 = g^{\mu\nu}R_{,\mu}R_{,\nu}$ \cite{Nozawa2008}.
In vacuum the mass function is constant on shell:
$\partial_\mu \mass = 0 \,  \rightarrow \, \mass = M =$ constant.
The most general vacuum solution with compact (positive curvature) horizon, 
given here in Schwarzschild-like coordinates, is
\bea
ds^2 &=& -F^2(R;M)dt^2 + F^{-2}(R;M)dR^2 + R^2 d\Omega \nonumber\\
 F^2(R;M) &=& 1+ \frac{R^2}{2\lambda_3}\left(1\mp
\sqrt{1+{4\lambda_3}\frac{2\kappa^2_5 M}{3R^{4}} -{4\lambda\lambda_3}} \right)
\ ,\label{eq:vacuum solution}
\eea
where $M$ is the on-shell value of the mass function and $F^2(R;M)=|DR|^2$ is 
obtained by inverting (\ref{eq:M}). The minus sign in front of the square root 
corresponds to the physical sector because it yields the 
Schwarzschild-Tangherlini-AdS solution in the limit that $\lambda_3\to 0$. 
Note that the GB term yields a modified effective cosmological constant
\be
\lambda_{eff} \equiv \left[ \frac{1}{2\lambda_3}\left(1-
\sqrt{1 -{4\lambda\lambda_3}}\right)\right]^{-1}\ ,
\label{eq:effective lambda}
\ee
as can be seen by taking either the $M\to 0$ or $R\to \infty$ limit in 
(\ref{eq:vacuum solution}).

The vacuum solution describes a single horizon black hole spacetime. In terms 
of the mass function, the horizon condition $|D R|^2_{R_H}=0$ is
\begin{equation}
\mathcal{M}(R_H)=\frac{1}{2}\left[\lambda R_H^{4} + R_H^{2}+{\lambda_3}\right]\ ,
\label{eq:massathorizon}
\end{equation}
which implies that $R_H\to 0$ as $\mathcal{M}(R_H)\to M_{crit}\equiv \lambda_3/2$
even in the dynamical context. This
suggests that it is impossible to form a black hole  when the ADM mass is 
less than this critical value. This is a special feature of 5D EGB,
as it depends critically on the exponent of $R_H$ in the third
term of the mass function.  It is similar to the existence of a critical
mass for black holes in AdS$_3$ with Einstein gravity.

\subsection{Hamiltonian Analysis}
%\subsection{Total Hamiltonian}

The total action describing the gravitational collapse of a massless scalar 
field in EGB gravity is
\be
I = I_{EGB} + I_M\ ,\ \textnormal{where}\ \
I_M = -\frac{1}{2}\int d^5x\sqrt{-g}\nabla_\mu\psi \nabla^\mu \psi\ .
\label{eq:TotalAction}
\ee
The Hamiltonian analysis of spherically symmetric EGB without cosmological 
constant was performed in \cite{Louko1997,Taves2012}
and extended to generic Einstein-Lanczos-Lovelock gravity in
\cite{Kunstatter2013}. 

Following \cite{Louko1997,Taves2012} we use the ADM parametrization
\be
ds^2 = - N^2dt^2 + \Lambda^2\left(dx+N_rdt\right)^2+ R^2d\Omega_{(3)}^2
\label{eq:adm metric}
\ee
and integrate out the angular coordinates in (\ref{eq:TotalAction}) 
to obtain a two dimensional dimensionally reduced action.
With this metric, we define
\be |DR|^2= -y^2 +\left(\frac{R'}{\Lambda}\right)^2\ \textnormal{where}\ \
y \equiv \frac{\dot{R}}{N}-\frac{N_r R'}{N} \ee
for future convenience.  Here and in the following, a dot is the partial
derivative with respect to $t$, and a prime is the partial with respect to
the radial coordinate $x$.

The dimensionally reduced Hamiltonian is a linear combination of the 
Hamiltonian constraint $\mathcal{G}$ and diffeomorphism constraint 
$\mathcal{F}$
\be
H_{tot} = \int dx \left(N\mathcal{G} + N_r\mathcal{F}\right)\ ,
\label{eq:Htot}
\ee
where we have dropped an overal factor equal to the integral over the 
unit three sphere. As shown in \cite{Louko1997},
\bea
\mathcal{G} &=& -\frac{6\lambda}{2\kappa_5^2} +  yP_R 
+ y^2\left[\Lambda R - \lambda_3 \left(\frac{R'}{\Lambda}\right)^\prime\right] 
-\Lambda R\left[1-\left(\frac{R'}{\Lambda}\right)^2\right]\nonumber \\
& &+\left(\frac{R'}{\Lambda}\right)^\prime\left\{R^2 + \lambda_3\left[1-\left(\frac{R'}{\Lambda}\right)^2\right]\right\} 
+ \frac{1}{2\Lambda}\left(\frac{P_\psi^2}{R^3}+ R^3 \psi'^2\right)\ ,\\
\mathcal{F} &=& R' P_R - \Lambda' P_\Lambda + \psi' P_\psi\ .
\eea
The momentum conjugate to $\Lambda$ is given by
\be
P_\Lambda = - \frac{3}{2\kappa_5^2}\left[ R^2 y  + 2\lambda_3 y\left(1- 
\left(\frac{R^\prime}{\Lambda}\right)^2\right)+2\lambda_3 \frac{y^3}{3}\right]
\ ,\label{eq:PLambda}
\ee
which determines $y=y(R,\Lambda,P_\Lambda)$ implicitly in terms of the other 
gravitational phase space variables. Note that $y$ is independent of $P_R$ 
and that we do not require the expression for $P_R$ in the following.

By taking suitable linear combinations of the Hamiltonian and 
diffeomorphism constraints, the total Hamiltonian (\ref{eq:Htot}) can, 
up to boundary terms, be written \cite{Kunstatter2013} as
\bea
H_{tot} &=&  \int dx \left[\left(\frac{N\Lambda}{R^\prime}\right)
\left(-\mass^\prime + \frac{R^\prime}{\Lambda^2}\rho_m-\frac{y}{\Lambda}P_\psi 
\psi^\prime \right) + \left(N_r+\frac{Ny}{R^\prime}\right)\mathcal{F} \right]
\label{eq:Htotb}\ ,
\eea
where 
\bea
\rho_m &=& \frac{1}{2} \left(\frac{P_\psi^2}{R^3}+ 
R^3\left(\psi^\prime\right)^2\right)
\eea
and  $\mass$ is the mass function  (\ref{eq:M}) expressed in terms of 
phase space variables. It is important for the following that $P_R$ appears 
only in the diffeomorphism constraint $\mathcal{F}$.

%\subsection{Choice of Coordinates}
We choose as spatial coordinate $R= R(x)$ with 
consistency condition $\dot{R}=0$, which requires
\be  
\frac{N_r}{N}=-\frac{y(R,\Lambda,P_\Lambda)}{R^\prime}\ .
\label{eq:Rdot}
\ee
We can now set the diffeomorphism constraint, gauge fixing condition, and 
consistency condition strongly to zero to obtain the partially reduced 
Lagrangian
 \be
L_{sr}(t) = \int dx \left(P_\Lambda \dot{\Lambda} +P_\psi \dot{\psi} 
- H_{sr}\right)\ ,
\label{eq:Lsr}
\ee
where
\be
H_{sr} =  \int dx \left(\frac{N\Lambda}{R^\prime}\right)
\left(-\mass^\prime + \frac{R^\prime}{\Lambda^2}\rho_m-\frac{y}{\Lambda}
P_\psi \psi^\prime \right)\ .
\ee

The remaining coordinate freedom can be fixed in two distinct ways. 
The first is to set the metric function $\Lambda=\Lambda(x)$ to be a 
specific function of $x$. We outline this procedure in an appendix.

The more common choice, namely Schwarzschild-like coordinates, is used 
for numerical studies in much of the current literature. 
This class is obtained by the choice $y(R,\Lambda,P_\Lambda) = 0$, 
which in turn implies that $P_\Lambda=0$. When $y=0$, (\ref{eq:M}) reduces to
\be
\mass= \frac{3}{2\kappa_n^2}R^{4}\left[\lambda + \frac{1}{R^2}
\left(1-\left(\frac{R^\prime}{\Lambda}\right)^2\right) 
+\frac{\lambda_3}{R^4}\left(1-\left(\frac{R^\prime}{\Lambda}\right)^2\right)^2
\right] \label{eq:M2}
\ee

%\subsubsection{Consistency Condition}
The consistency condition, $\dot{y}=0$, for this gauge choice is
\be
\left(\frac{N\Lambda}{R^\prime}\right)^\prime=-\frac{\kappa^2_5}{3}
\left(\frac{N\Lambda}{R^\prime}\right)\frac{R}{R^\prime} 
\frac{\Pi^2+\Phi^2}{\left[\frac{1}{R^2}+ 2 \frac{\lambda_3}{R^4}
\left(1-\left(\frac{R^\prime}{\Lambda}\right)^2\right)\right]}\ ,
\label{eq:Nconsistency}
\ee
where
\be
\Pi\equiv \frac{P_\psi}{R^{n-2}}\ ,\qquad \Phi \equiv \psi^\prime\ .
\ee
Using the Hamiltonian constraint,
\be
\mass^\prime = \frac{R^\prime}{\Lambda^2}\rho_m 
=  \frac{R^{n-2}R^\prime}{2\Lambda^2}(\Pi^2+\Phi^2)\ .
\label{eq:ham2}
\ee
We note that in vacuum $(N\Lambda/ R')' =0$ and the constraint $\mass'=0$  
can be solved algebraically for $\Lambda$ and $N$ to give 
(\ref{eq:vacuum solution}).

The dynamical equations can be obtained by varying the following fully 
reduced Hamiltonian with respect to $\psi$ and $P_\psi$ ($\Pi$ and 
$\Phi$ are not canonical variables):
\be
H_{red}=\int dx \left(\frac{N\Lambda}{R'}\right)\left[-\mass^\prime+
\frac{R^\prime}{2\Lambda^2}R^{n-2}\left(\Pi^2+ \Phi^2\right)\right] \ .
\label{eq:Hred}
\ee
In the above, $N$ and $\Lambda$ are implicitly defined by the consistency 
condition (\ref{eq:Nconsistency}) and Hamiltonian constraint (\ref{eq:ham2}), 
respectively. They do not need to be varied, however, since the corresponding 
variations of $H_{red}$ are proportional to the Hamiltonian constraint 
and consistency condition.
The resulting evolution equations are
\be
\dot{\Phi} = \left(\frac{N}{\Lambda}\Pi\right)^\prime\ \textnormal{and}\ 
\dot{\Pi} =  \frac{1}{R^3}\left(\frac{ N}{\Lambda} R^{3}\Phi\right)^\prime\ .
\ee
The above, along with (\ref{eq:M2}), (\ref{eq:Nconsistency}), 
and (\ref{eq:ham2}) are the complete set of equations to solve.

We now put these equations into the form used in \cite{Deppe:2014oua} 
by making the substitutions
\be
\Lambda^2 = \frac{ R^\prime}{A}\ ,\ N^2 =  R^\prime A e^{-2\delta}\ 
\Rightarrow\ \frac{N}{\Lambda}=Ae^{-\delta}\ .\ee
We choose a compactified spatial coordinate $R(x) = l \tan(x/l)$
with $l= 1/\lambda_{eff}$.
The metric in terms of dimensionless coordinates $x\to x/l$, $t\to t/l$ is
\be
ds^2 = \frac{1}{\cos^2(x)}\left(-Ae^{-2\delta} dt^2 + A^{-1}dx^2 + \sin^2(x) 
d\Omega^{(n-2)}\right)\ ,
\ee
while the Hamiltonian constraint in terms of the new metric functions becomes
\be
\mass
=  \frac{3}{2\kappa_n^2}R^{4}\left[ \frac{\lambda_2}{l^2\sin^2(x)}(1-A) +
   \frac{\lambda_3}{l^4\sin^4(x)}(1-A)^2\right]\ ,
\label{eq:M2b}
\ee
where $\lambda_2= 1- 2\lambda_3/l^2$ and we have used the identity
\be
\frac{1}{R^2}\left(1- {R^\prime} A\right)= -1 + \frac{1}{\sin^2(x)}(1-A)\ .
\ee
We make the scalar field and its conjugate dimensionless by rescaling
$\Phi \to \kappa_5\Phi/\sqrt 3$ and $\Pi\to \kappa_5\Pi/\sqrt 3$.
Finally, we absorb $l^2$ into the mass function and $\lambda_3$ to make them 
dimensionless as well.

In the end, we solve the following equations:
%%%%%%%%%%%%%%%%%%
\bea
\dot\Phi%_{,t}
&=& \left(Ae^{-\delta}\Pi\right)'\\ %_{,x}\\
\dot\Pi%_{,t}
&=& \frac{3}{\sin(x)\cos(x)}Ae^{-\delta} \Phi + \left(Ae^{-\delta}\Phi
\right)'\\ %_{,x}\\
\delta'%_{,x}
&=& - \frac{\cos(x)\sin^3(x) (\Pi^2+\Phi^2)}{\left[\sin^2(x) - 
2 \lambda_3\left(A-\cos^2(x)\right)\right]}
\label{eq:deltaPrime2}\\
\mass'%_{,x}
&=& \frac{A}{2}\tan^{3}(x) (\Pi^2+\Phi^2)
\label{eq:Mprime2}\\
A &=&1+ \frac{\sin^2(x) (1-2\lambda_3)}{2\lambda_3}\left[1 -  \sqrt{1 + 
\frac{8\mass\lambda_3}{(1-2\lambda_3)^2\tan^{4}(x)}}\right]
\label{eq:Afinal}
\eea

%%%%%%%%%%%%%%%%%%%%%%%%%%%%%%%%%%%%%%%%%%%%%
Since nonlinear self-gravitation effects drop off sufficiently quickly at
large radius due to the dilution of energy density, 
the scalar field satisfies the same asymptotic expansion as
in the linearized theory,
$\Phi = \rho^3\left(\Phi_0 + \Phi_2\rho^2+\cdots\right)$ and  
$\Pi  = \rho^2\left(\Pi_0 + \Pi_2 \rho^2 + \cdots\right)$, where
$\rho=\pi/2-x$.  These are the boundary conditions of the 
normalizable linear eigenmodes $e_k(x)$, which can be defined in terms of
Jacobi polynomials; the leading terms in these expansions correspond to
expectation values of operators in the dual field theory.\footnote{There are 
also non-normalizable scalar modes (ignoring gravity) which lead to a 
different asymptotic expansion and correspond to spacetime-varying operators
in the Hamiltonian of the dual theory.}  At the origin, we require that
$\Pi$ be an even function of $x$ and $\Phi$ be odd for smoothness.

\subsection{Numerical Methods}\label{s:numerical methods}

We  briefly outline  our  numerical  methods  and  how  these  have  changed  
since  our  previous work \cite{Deppe:2014oua}.  A detailed description is 
provided in an appendix of \cite{Deppe:2015qsa}.  The key improvement to
our code is that we now solve the spatial ordinary differential equations 
using an adaptive fifth-order Dormand-Prince stepper.  We set the desired 
relative and absolute tolerances and the stepper will adjust the step size 
over the spatial mesh so that the desired tolerances are met locally.  
The adaptive method requires scalar field data in between grid points which we
supply using a cubic spline.  We find that the stepper takes many small steps 
near the origin and much larger steps further out.

\section{Results}\label{s:results}

\begin{figure}[t]
\centering
\includegraphics[width=0.70\textwidth]{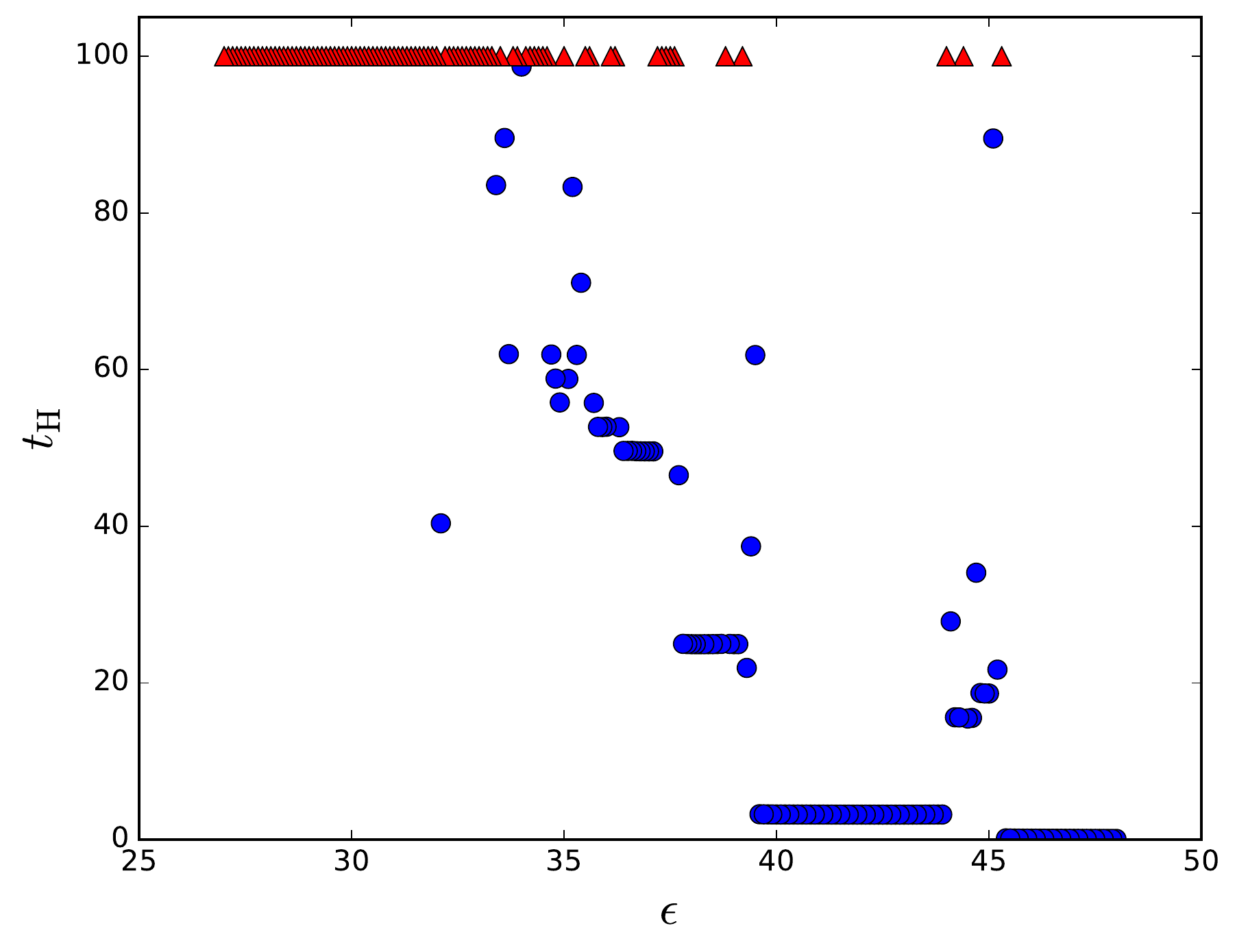}
\caption{Horizon formation times $t_H$ as a function of amplitude $\epsilon$.
Blue circles represent amplitudes which form a horizon; red triangles
represent lower limits for $t_H$ based on simulations at fixed resolution.}
\label{fig:GBscan}
\end{figure}

As in \cite{Deppe:2014oua}, we consider initial data of the form
\begin{equation}
\phi=\Phi=0\ ,\ \  \Pi = \frac{2}{\pi}\epsilon\exp\left(-\left(
\frac{2}{\pi}\frac{\tan(x)}{\sigma}\right)^2\right)\ ,\ \ 
\sigma=\frac{1}{16}
\end{equation}
ie, Gaussian in $\Pi$, and a GB parameter of $\lambda=0.002$.  
Figure \ref{fig:GBscan} provides an overview of our results for the horizon
formation time $t_H$, which cover an amplitude range $\epsilon=27-48$.
In the figure, blue circles represent formation of a horizon, while red
triangles represent lower limits on $t_H$ for amplitudes which do not form
a horizon for $t<100$.  For Einstein gravity ($\lambda=0$), $t_H$ would be
approximately piecewise constant appearing as ``steps'' with $t_H$ with 
decreasing amplitude.  Physically speaking, at large amplitude, gravitational
collapse can proceed immediately, but lower amplitude initial data disperses,
reflects from the conformal boundary one or more times, and finally collapses
after more gravitational focusing.

As in the earlier results of \cite{Deppe:2014oua}, gravitational collapse
in EGB gravity exhibits the same as well as additional features.  First,
there is a transition from immediate collapse to collapse after one or more
reflections.  There is critical behavior at these transitions, which has
been studied in some detail in \cite{Olivan:2015fmy,Santos-Olivan:2016djn}
for Einstein gravity in AdS.  In the 
following subsection,  we study the first critical point at large amplitude,
when horizon formation stops occurring immediately, extending the analysis
of this region in \cite{Deppe:2014oua}.  

Another key feature of figure \ref{fig:GBscan} is that $t_H$ appears to
demonstrate sensitivity to initial conditions in certain amplitude ranges.
That is, while there are some steps in horizon formation time where
$t_H$ remains approximately constant with $\epsilon$, $t_H$ varies wildly
in transition regions between the steps.  At low enough amplitude, the 
steps are apparently so narrow that they dissolve into the transition regions.
In subsection \ref{s:chaotic}, we explore in more detail whether the
transition regions exhibit chaotic behavior such as self-similarity.

Because horizon formation is apparently sensitive to initial conditions in
some regions of the amplitude, we have opted to keep all
the simulations of figure \ref{fig:GBscan} at a fixed
resolution of $2^{12}+1$ grid points, even when they begin to lose convergence
(as illustrated by a loss of conserved ADM mass).  Otherwise, an increase
in resolution could act as a small shift in amplitude.
  At this resolution, simulations lose up to 2.5\% 
of the conserved ADM mass by $t=100$, so simulations that do not form a 
horizon by this time are shown only as lower limits on $t_H$.
We have tested several amplitudes with $2^{13}+1$ grid points and found that
subcritical simulations remain subcritical while horizon formation times 
in the step regions (which are stable vs change of initial conditions) have
a relative difference of $5\times 10^{-7}$.

Finally, at the lowest amplitudes shown, none of the simulations form a 
horizon.  As noted earlier, horizons cannot form below a critical conserved
mass $M_{crit}$ in EGB gravity. 
In other words, all initial data must be stable at low amplitudes, 
in apparent contrast to the case of Einstein gravity.  For our choice of 
initial data, the critical mass corresponds to an amplitude of approximately 
$\epsilon_{crit}\sim 21.86$; figure \ref{fig:GBscan} hints that higher 
amplitudes may also be dynamically stable against horizon formation.  
It is also an interesting question whether evolution of initial data below
the critical amplitude is quasi-periodic or evolves toward a naked singularity.
In section \ref{s:naked}, we study the evolution for two amplitudes, one just
larger than and one just smaller than the critical amplitude, and present
evidence suggestive of naked singularity formation at finite origin time
below the critical amplitude.

\subsection{Critical Phenomenon}\label{s:critical phenomenon}
Critical phenomena in the gravitational collapse of a spherically
symmetric scalar field in Einstein gravity (for 
4-dimensional asymptotically flat spacetime)
was first observed by Choptuik \cite{Choptuik:1992jv}. Choptuik found that 
geometrical quantities such as the mass of the black hole obey the scaling law
\begin{align}
  \label{eq:choptuikRelation}
  M_{BH}\propto(p-p_\star)^\gamma
\end{align}
where $p$ is a parameter in the initial data profile, $p_\star$ is the
critical value of $p$,\footnote{For $p>p_\star$ the scalar field collapses to
a black hole, and for $p<p_\star$ the field disperses.}, and $\gamma$ is the
critical exponent. A detailed semi-analytic study by Gundlach in four 
dimensions \cite{Gundlach:1996eg} found that $\gamma=0.374\pm 0.001$. 
For small amplitude initial data in asymptotically AdS spacetime, 
any horizon that forms will be small compared
to the AdS scale, so the critical behavior at any transition (ie, collapse
after $n$ reflections transitioning to collapse after $n+1$ reflections)
should have the same critical exponent as the asymptotically flat case.
In the case of Einstein gravity, \cite{Santos-Olivan:2016djn} confirms
the expectation, finding a critical exponent consistent with the Gundlach
value independent of the width of initial data or the number of reflections
before collapse.

The critical behavior of EGB gravity differs from Einstein gravity even
in asymptotically flat spacetime.  For one, the Gauss-Bonnet term contributes
to the equations of motion only in 5 dimensions or more; in 5D Einstein
gravity, the critical exponent is $\gamma\approx 0.416$
\cite{Bland:2005kk,Taves:2011yt}.  
Critical phenomena in 5D EGB gravity has been studied in
\cite{Golod:2012yt,Deppe:2012wk}, which found that the new, short distance 
length scale alters the near-critical behavior
such that no black hole forms below a minimum horizon radius 
\cite{Deppe:2012wk}.  This is similar to the case of a massive scalar
field in asymptotically flat spacetime, which also has a dynamically 
determined minimum horizon radius \cite{Brady:1997fj}.

\begin{figure}[t]
  \centering
  \includegraphics[width=0.50\textwidth]{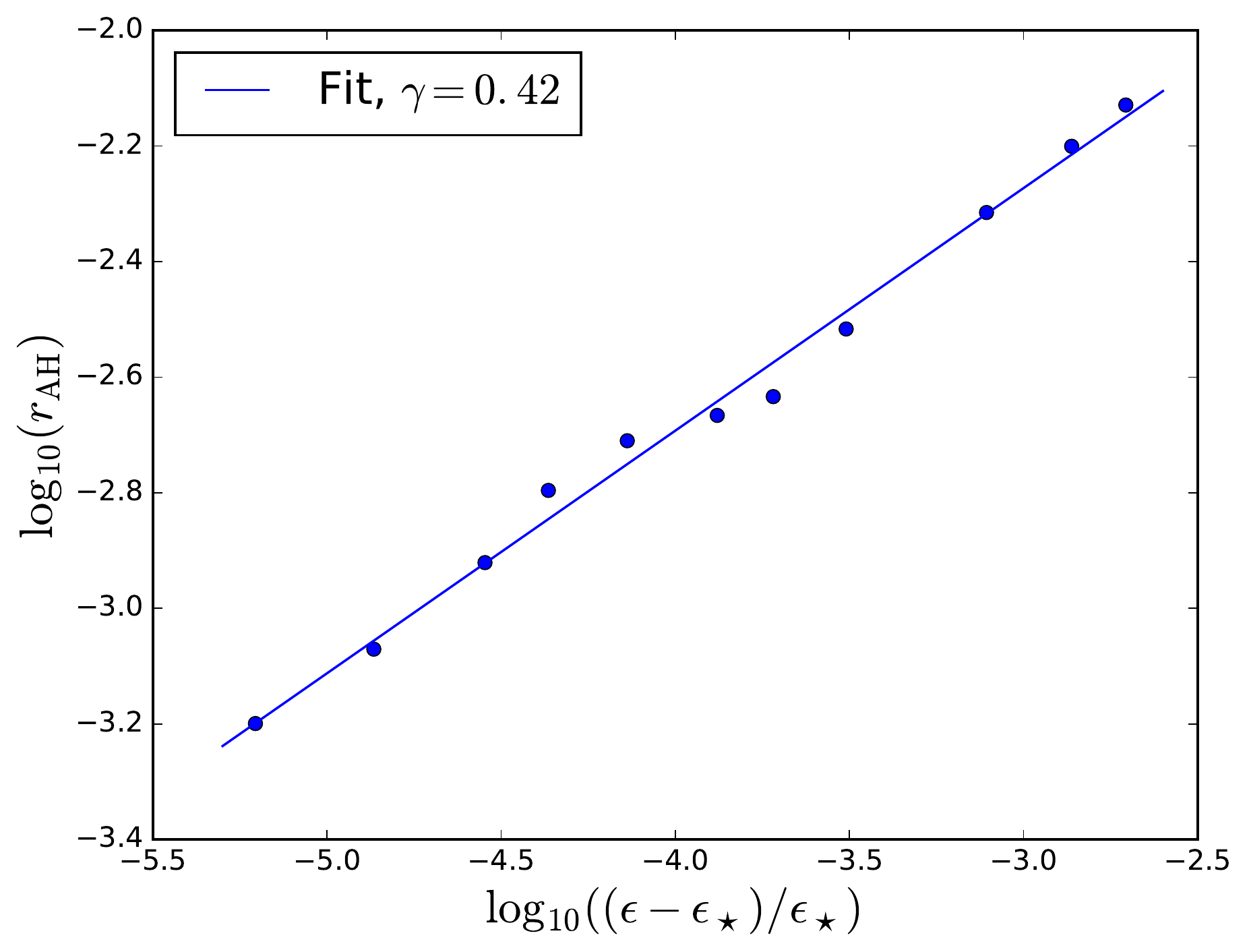}
  \caption{Initial horizon radius vs amplitude for the immediate collapse
    for Einstein gravity. ($\gamma\approx0.42$)  }
  \label{fig:GRScaling}
\end{figure}

Again, it is natural to ask which features of the critical behavior persist
or differ in asymptotically AdS$_5$ spacetime.  As in 4D, we expect 
critical behavior near each transition ($n$ to $n+1$ reflections) to 
match that in asymptotically flat spacetime because the black holes formed
are initially much smaller than the AdS curvature scale.
In figure \ref{fig:GRScaling}, we show
$\log(r_{\rm{AH}})$ as a function of 
$\log((\epsilon-\epsilon_\star)/\epsilon_\star)$, where $\epsilon_\star$ 
is the amplitude above which scalar field configurations collapse
immediately (the 0 to 1 reflection transition).  
We find a critical exponent  $\gamma\approx 0.42$ in agreement
with results in asymptotically flat spacetime.

\comment{
A natural question to ask is if the addition of a negative
cosmological constant alters this behavior, since now the matter
cannot disperse and is bounded. Near the critical solution
only some of the scalar
field collapses to form the black hole, with the rest
dispersing. However, since the field is now bounded, it can reflect
and fall into the black hole at a later time. If evolutions are run
sufficiently long
then this should be observable as a jump in the radius of the black
hole. However, in practice simulations are usually stopped
within a time of $t\approx0.7$ for the nearest to $\epsilon_\star$
evolutions. This is much less than the transit time for the scalar
field to reflect off the boundary and return to the origin, which is
approximately $\Delta t=\pi$. }

\begin{figure}[t]
  \begin{subfigure}{0.48\textwidth}
    \includegraphics[width=0.95\textwidth]{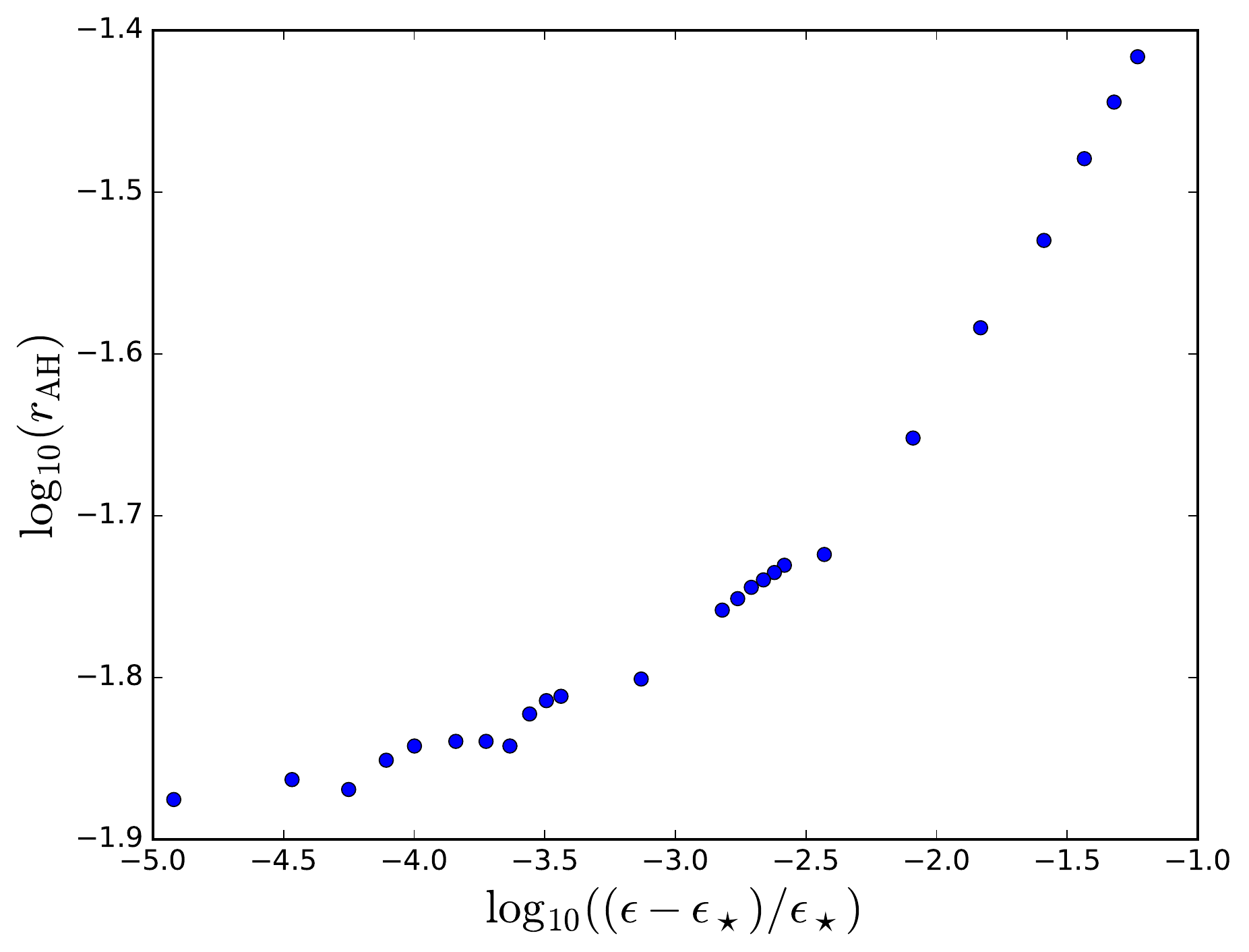}
    \caption{Evidence for the radius gap}
    \label{fig:RadiusGap}
  \end{subfigure}
  \begin{subfigure}{0.48\textwidth}
    \includegraphics[width=0.95\textwidth]{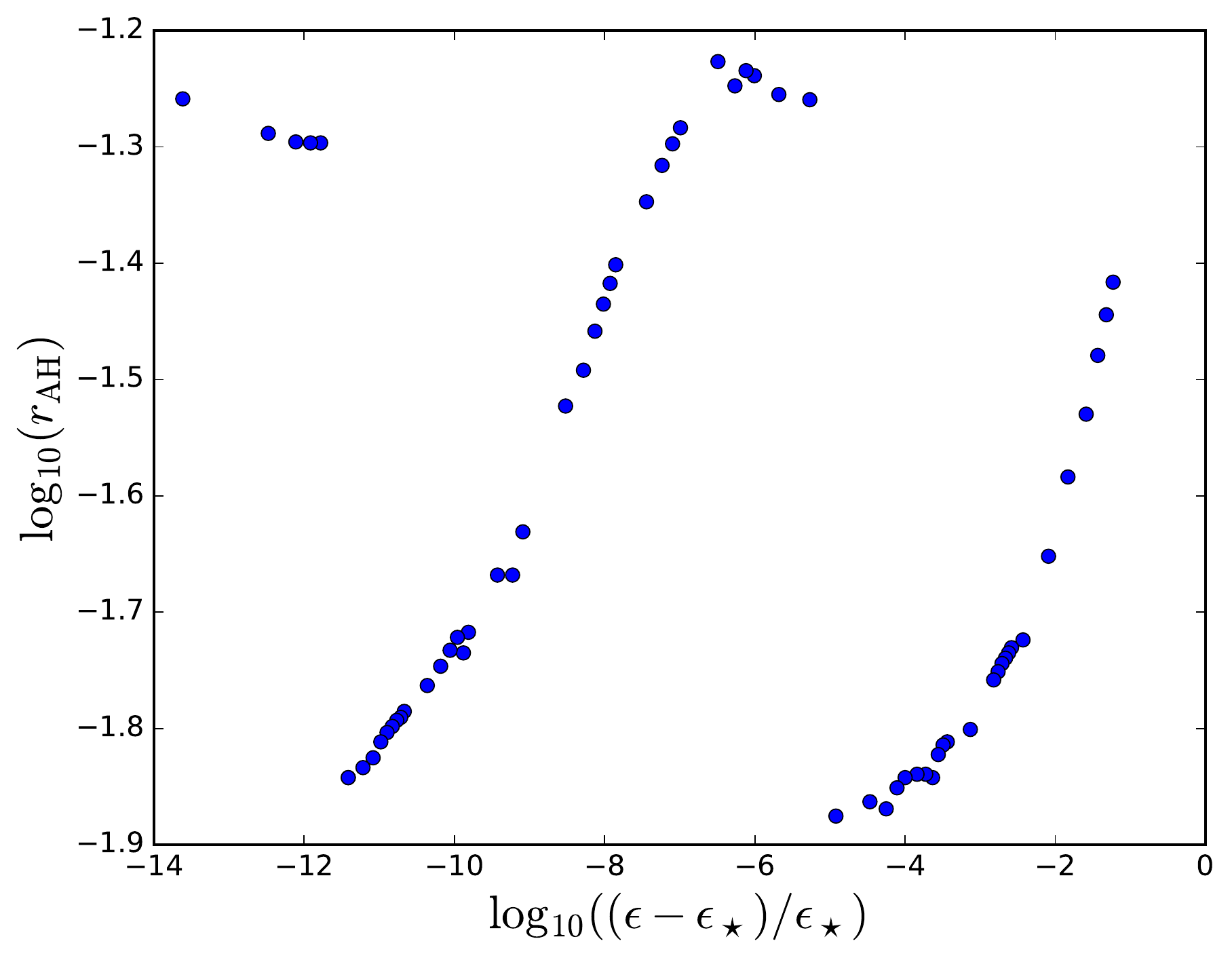}
    \caption{Step-like behavior from time dilation}
    \label{fig:StepsInScaling}
  \end{subfigure}
  \caption{Initial apparent horizon radius vs amplitude near the critical
    point for immediate collapse in 5D EGB gravity.}
  \label{fig:GBchoptuik}
\end{figure}

For EGB gravity, we expect a minimum horizon radius at each critical 
point, as in asymptotically flat spacetime.  Figure \ref{fig:GBchoptuik}
shows the scaling of the initial apparent horizon radius with
amplitude near the critical point for immediate collapse, 
$\epsilon_\star\approx 45.3315$.  It is initially apparent 
from figure  \ref{fig:RadiusGap}, which shows values of 
$\epsilon$ far from $\epsilon_\star$ where the black holes form very quickly, 
that there is in fact a radius gap as seen in \cite{Deppe:2014oua}.
Continuing to amplitudes with 
$\epsilon-\epsilon_\star\lesssim 10^{-5}\epsilon_\star$ in figure
\ref{fig:StepsInScaling}, we observe persistence of the radius gap
$R_{min}\sim 10^{-1.9}$
along with sudden jumps, or steps, in the horizon radius.

\begin{figure}[t]
  \centering
  \includegraphics[width=0.70\textwidth]{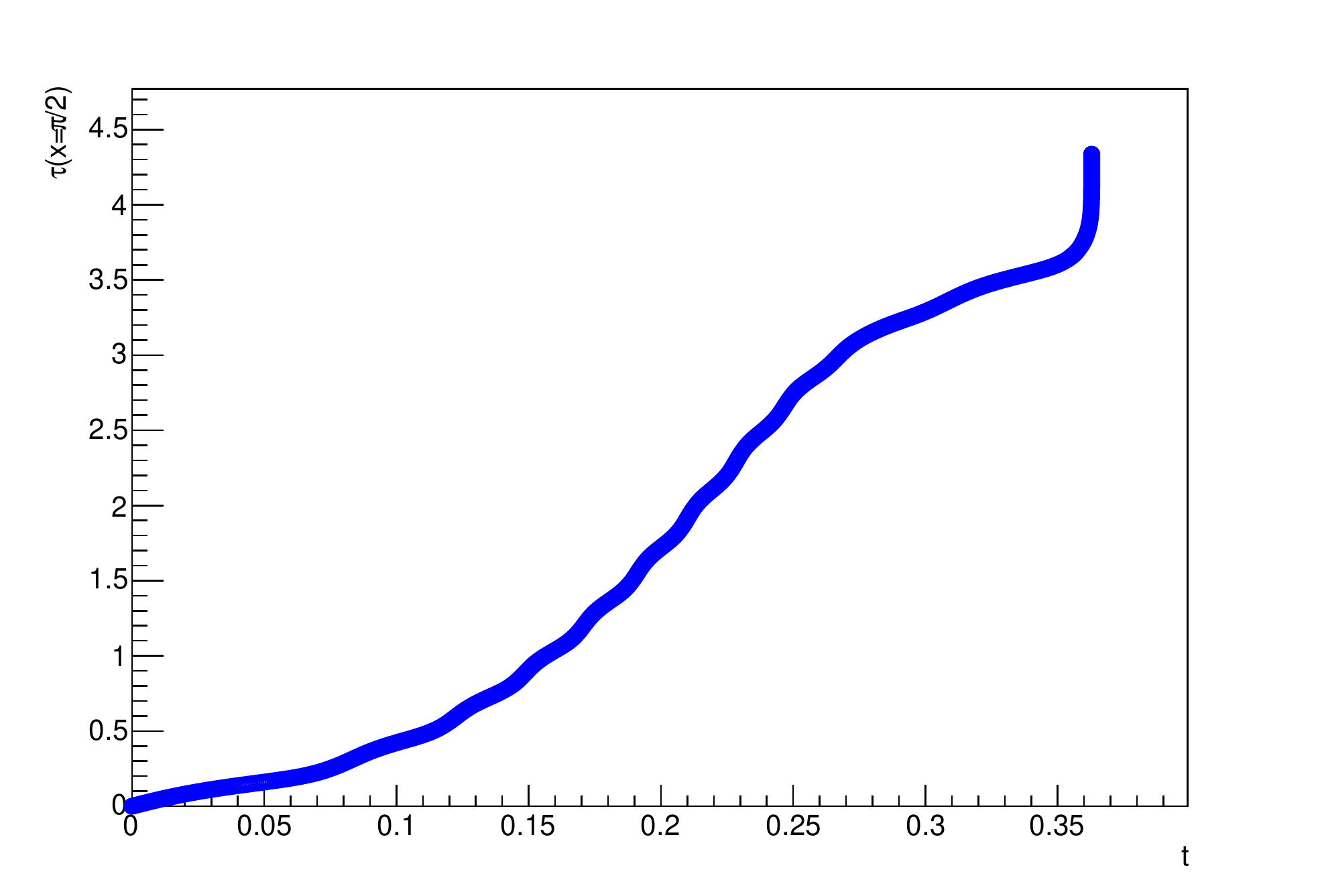}
  \caption{Plot of the proper time at the AdS boundary as function of
    the proper time of an observer at the origin. The strong redshift
    effects allows the dispersed matter to reflect of the AdS boundary
    and interact with the black hole, causing an abrupt jump in the
    radius of the black hole.}
  \label{fig:TimeDilation}
\end{figure}

An explanation for this behavior is apparent in animations of our
simulations.  As the scalar field collapses, the initial profile fragments,
with the main portion of the mass remaining near the origin and driving
horizon formation while several pulses of mass disperse toward the 
boundary.  For $\epsilon-\epsilon_\star$ small, one or even two of these
subsidiary pulses can reflect from the boundary and return to the 
neighborhood of the origin (possibly multiple times) 
before $A(t,x)$ reaches the threshold for 
approximate horizon formation.  These subsidiary pulses are responsible
for the multiple local minima of $A(t,x)$ noted in \cite{Deppe:2014oua}.
Animations showing sub-pulses reflecting from the boundary once and twice are
available at \url{http://ion.uwinnipeg.ca/~afrey/AdSGB16.html}.
Although the horizon formation times $t_H$ for these amplitudes are small,
it is important to remember that $t$ is the proper time at the origin.
As it turns out, there is a significant redshift factor between this time
and the proper time outside $x\gtrsim r_{AH}$.  To quantify the time
dilation factor, in figure \ref{fig:TimeDilation}
we plot the proper time of an observer at the AdS boundary, given by
\begin{align}
  \label{eq:boundaryTime}
  \tau=\int_0^t\exp\left[-\delta(t,x=\pi/2)\right]\,dt,
\end{align}
as a function of the proper time at the origin (in one particular
collapse).  While
an insufficient amount of time apparently passes for the scalar field to
reflect off of the AdS boundary according to observers at the origin, 
the relevant time is actually better approximated by the proper time at the 
AdS boundary since $\delta(t,x)$ is roughly spatially constant outside the
main portion of matter. Specifically,
while only a time $\Delta t\approx 0.37$ passes, the corresponding
boundary time elapsed is $\Delta \tau\approx 4.4$, enough for the
subsidiary pulses to reflect off the boundary and interact with the forming
black hole.  Interestingly, this effect should be observable in Einstein
gravity close enough to the critical amplitude (since infinite boundary 
time passes before the metric function $A(t,x)\to 0$), but it appears
to be much more challenging to resolve. Some progress on the subject
has been made \cite{Allahyari:2014lta}, but a different gauge choice
and black hole excision may be needed to fully explore this
behavior.  The GB term seems to enhance time dilation effects significantly.

\subsection{Self-Similarity}\label{s:chaotic}

As we noted above, the initial horizon formation time $t_H$ exhibits a much
richer structure in EGB gravity than in pure Einstein gravity.  By now it
is well-known that $t_H$ typically increases piecewise monotonically with
decreasing $\epsilon$ for massless scalar matter in Einstein gravity,
forming the ubiquitous step structure seen in many references.\footnote{There
are some widths $\sigma$ for the initial data on the ``coastlines'' of 
islands of stability with non-monotonic $t_H$
\cite{Buchel:2013uba,Deppe:2015qsa}.}
In contrast, while figure \ref{fig:GBscan} also exhibits some steps in $t_H$,
the transitions from step to step exhibit a significant non-monotonic
scatter in $t_H$.  For example, while initial data with $\epsilon\gtrsim 45.3$
collapses immediately and initial data with 
$39.6\lesssim \epsilon\lesssim 44.0$ collapse after one reflection from the
boundary, amplitudes $44.0\lesssim\epsilon\lesssim 45.3$ vary wildly.
The appearance of smaller steps and apparent sensitivity
to initial conditions in the transition regions led \cite{Deppe:2014oua} 
to speculate that the $t_H$ vs $\epsilon$ curve may have a fractal structure.
Here, we investigate the transition region $44.0\leq\epsilon\leq 45.3$
in more detail with the aim of uncovering signatures of chaotic behavior.
Since changing resolution amounts to a change in initial conditions, all 
simulations discussed in this subsection are carried out at a fixed resolution
of $2^{14}+1$ grid points, following the same reasoning explained above.

\begin{figure}[t]
\begin{subfigure}{0.48\textwidth}
\includegraphics[width=0.95\textwidth]{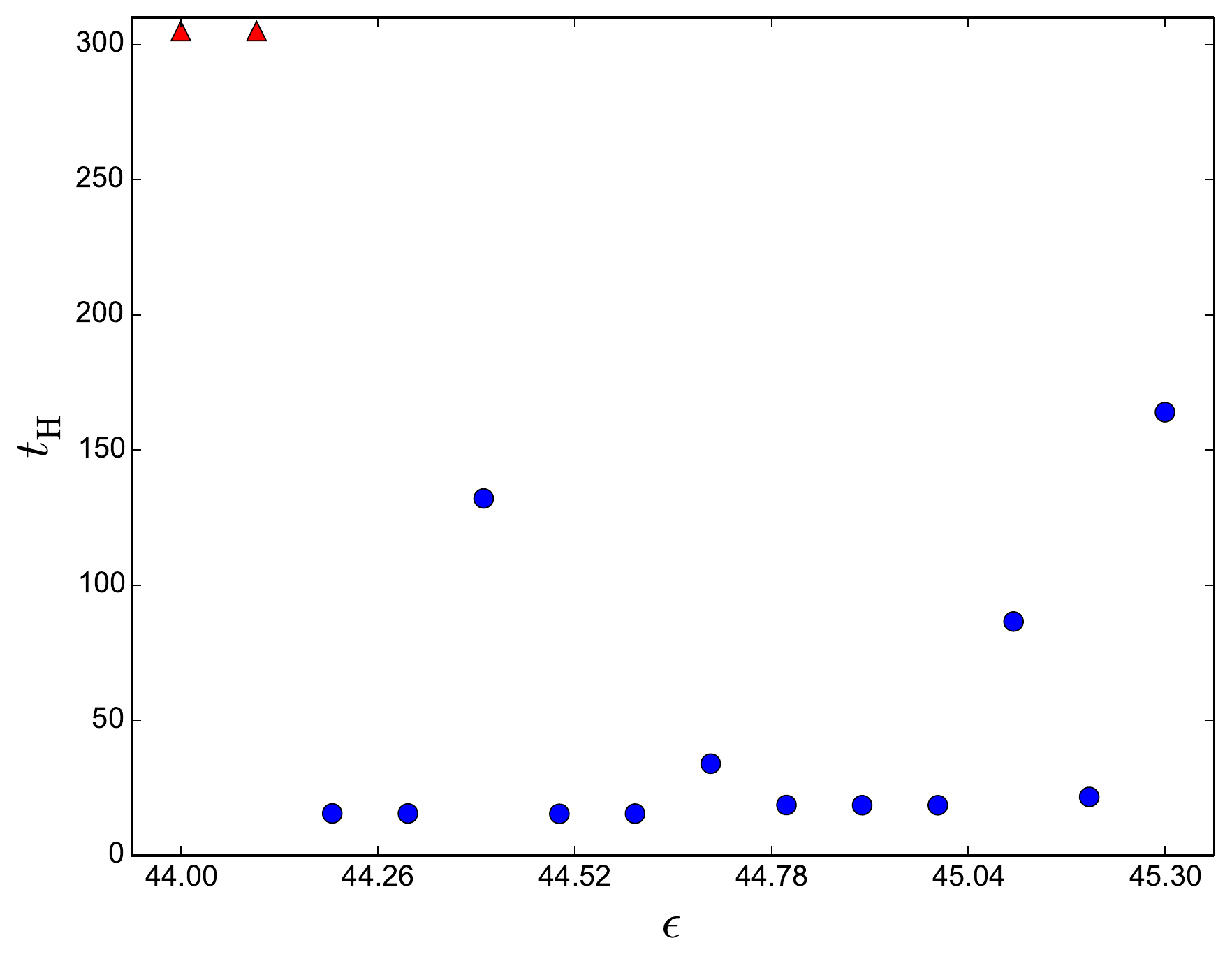}
\caption{$\epsilon=44.0$ to $45.3$}
\label{fig:chaosZoom1}
\end{subfigure}
\begin{subfigure}{0.48\textwidth}
\includegraphics[width=0.95\textwidth]{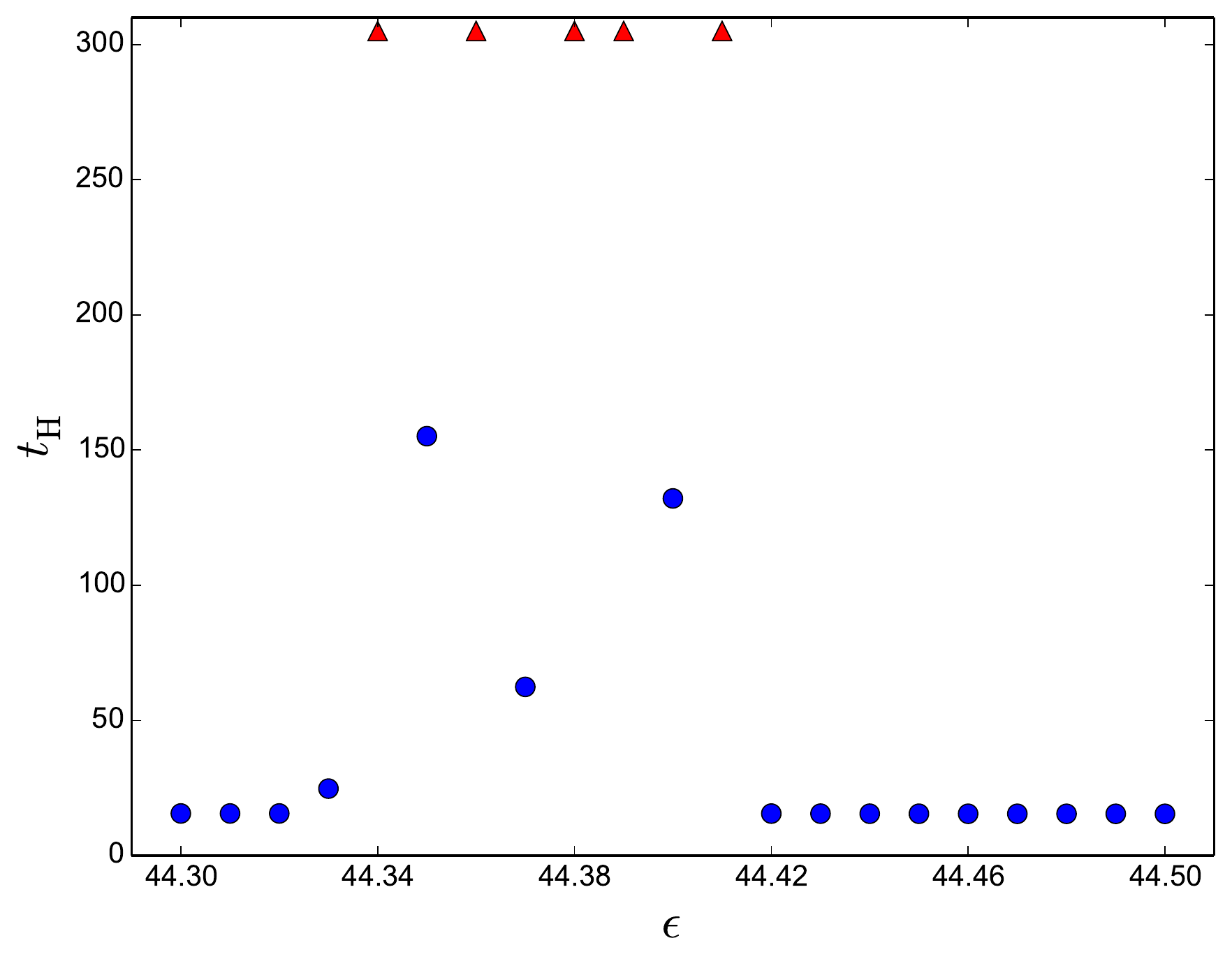}
\caption{$\epsilon=44.3$ to $44.5$}
\label{fig:chaosZoom2}
\end{subfigure}
\begin{center}\begin{subfigure}{0.48\textwidth}
\includegraphics[width=0.95\textwidth]{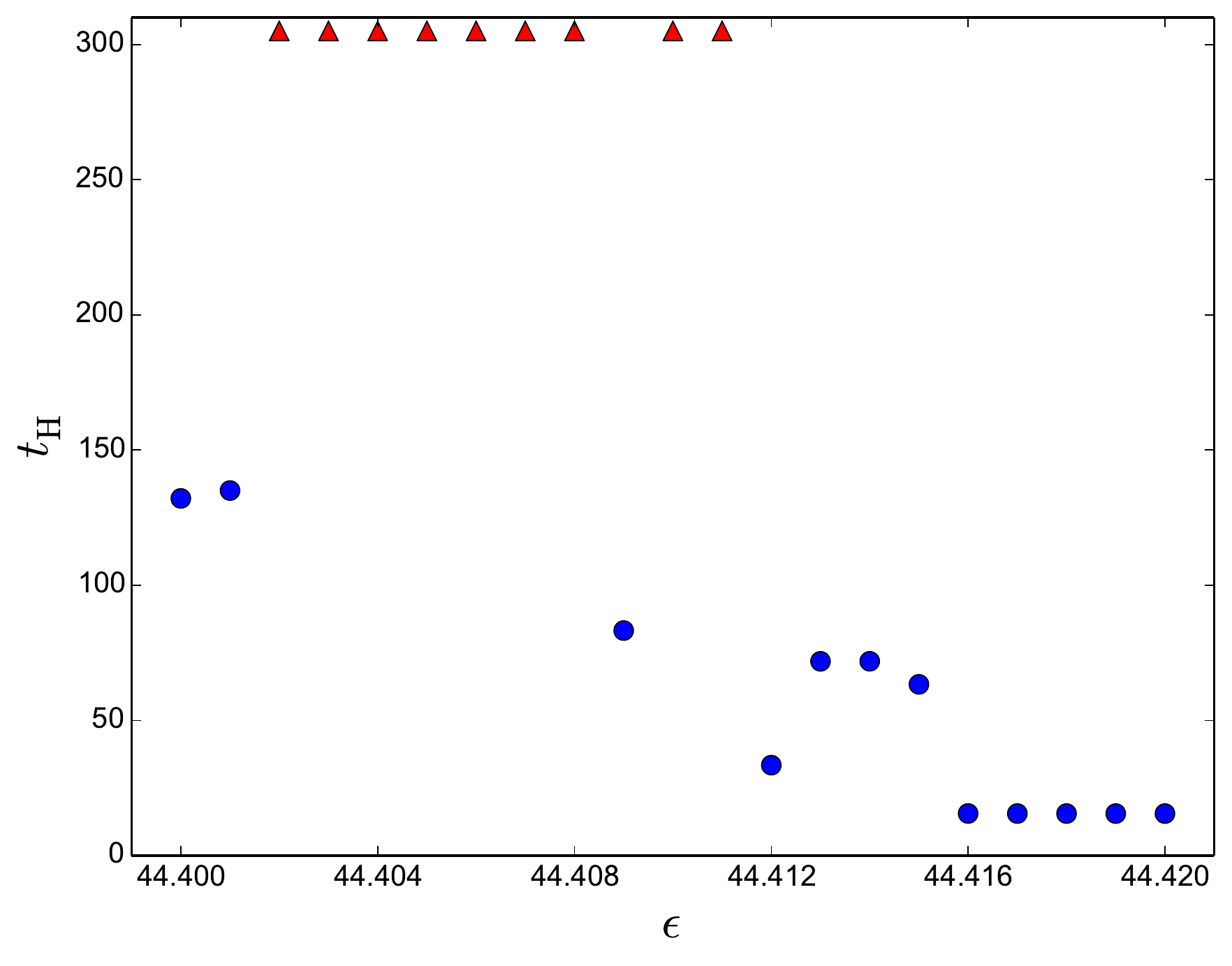}
\caption{$\epsilon=44.40$ to $44.42$}
\label{fig:chaosZoom3}
\end{subfigure}\end{center}
\caption{$t_H$ vs amplitude in the transition region from immediate collapse
to collapse after $1$ reflection, $\epsilon\approx 44.0$ to $45.3$, over
several ranges.}
\label{fig:chaosPlots}
\end{figure}

Figure \ref{fig:chaosPlots} shows the transition region in detail for three
ranges; figure \ref{fig:chaosZoom1} shows the entire region, figure
\ref{fig:chaosZoom2} shows a small area surrounding the $t_H\approx 132$
point at $\epsilon=44.4$, and \ref{fig:chaosZoom3} shows 
a smaller area to the right of that point.
Blue circles represent horizon formation, while red triangles represent
simulations that do not form a horizon within $t=305$, which can be taken 
as a lower limit on $t_H$ for those amplitudes.  These simulations 
lose several percent of the ADM mass by that time, however, so a conservative 
reader may prefer to read these as lower limits of $t_H\gtrsim 170$,
just greater than the largest values of $t_H$ for collapsing simulations.
Regardless, the plots for the three amplitude ranges show a similar structure
of rapidly varying horizon formation times with amplitude.  This remains
suggestive of fractal-like, self-similar behavior, at least on the scales
shown.

To test the self-similarity of the $t_H$ vs $\epsilon$ curve quantitatively,
we use a variation on the box-counting-dimension estimate.  Specifically,
we draw grid lines at each of the tick marks on figure \ref{fig:chaosZoom1}
and count the number of boxes so created that are occupied by data points.
For a first estimate, we include subcritical simulations as if they have
$t_H=300$.  Data points lying on a grid line are counted as occupying the
box above or to the right as appropriate.  In this case, a curve of dimension
$D$ should occupy $N=W/s^D$ boxes, where $W$ is the total horizontal range
($W=1.3$ in figure \ref{fig:chaosZoom1}) and $s$ is the length between
grid lines ($s=0.26$ in figure \ref{fig:chaosZoom1}); the box-counting
dimension is defined as $D=-\lim_{s\to 0} \ln(N/W)/\ln s$.  To take an 
approximate limit, we repeat the procedure for figures \ref{fig:chaosZoom2}
and \ref{fig:chaosZoom3}, keeping vertical grid lines at the tick marks shown
but scaling the vertical distance between horizontal grid lines with $s$.
We find $N=9,10,11$ and consequently $D_s=1.44,1.22,1.14$ respectively for
the three subfigures ($s=0.26,0.04,0.004$).   
However, it is reasonable to argue that the 
amplitudes that do not form a horizon may have different values of $t_H$ 
from each other (or be truly stable),
so we should not count them.  If we repeat the box-counting test while
ignoring the apparently stable points and also subtracting from $W$ the
width of any boxes that contain no collapsing data points, we find
$D_s=1.35,1.15,1.13$ for the three subfigures.  This provides weak but 
suggestive evidence that the $t_H$ vs $\epsilon$ curve has a fractal dimension
of around 1.14, somewhat greater than unity.

\begin{figure}[t]
\begin{subfigure}{0.48\textwidth}
\includegraphics[width=0.95\textwidth]{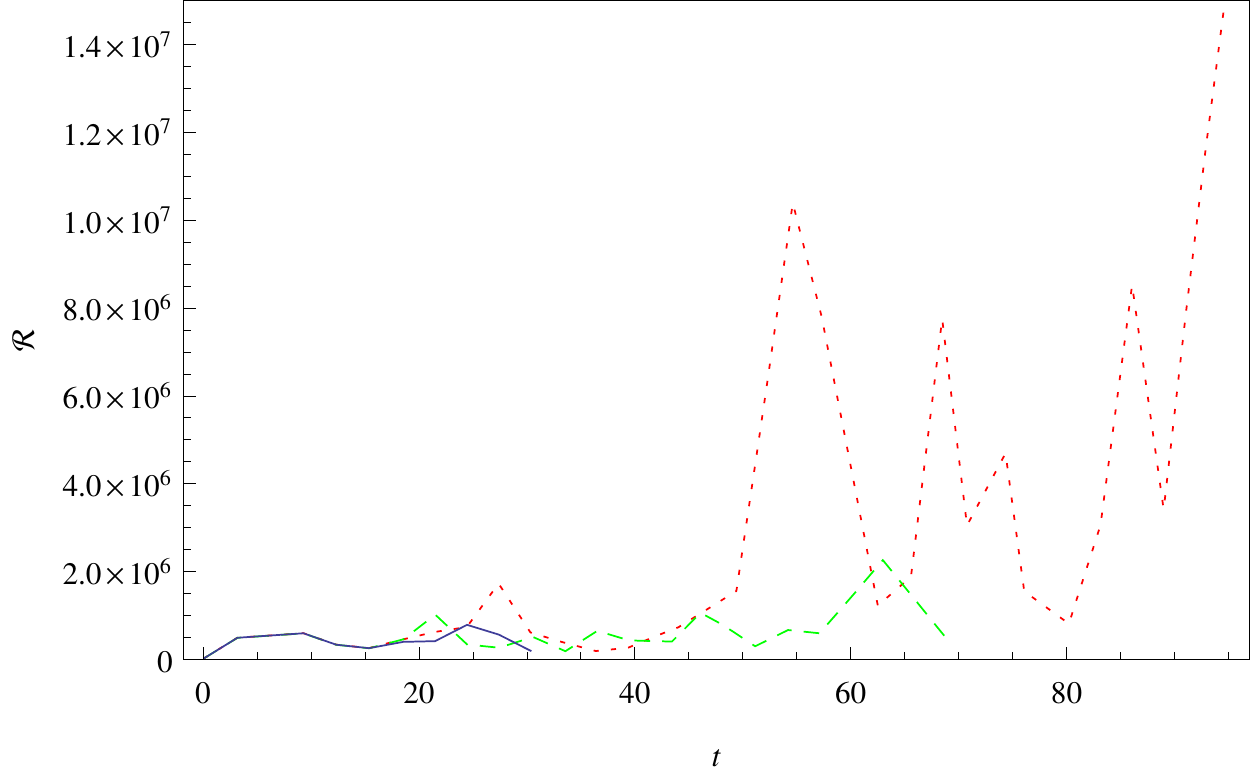}
\caption{Ricci scalar at origin}
\label{fig:lyapRicci}
\end{subfigure}
\begin{subfigure}{0.48\textwidth}
\includegraphics[width=0.85\textwidth]{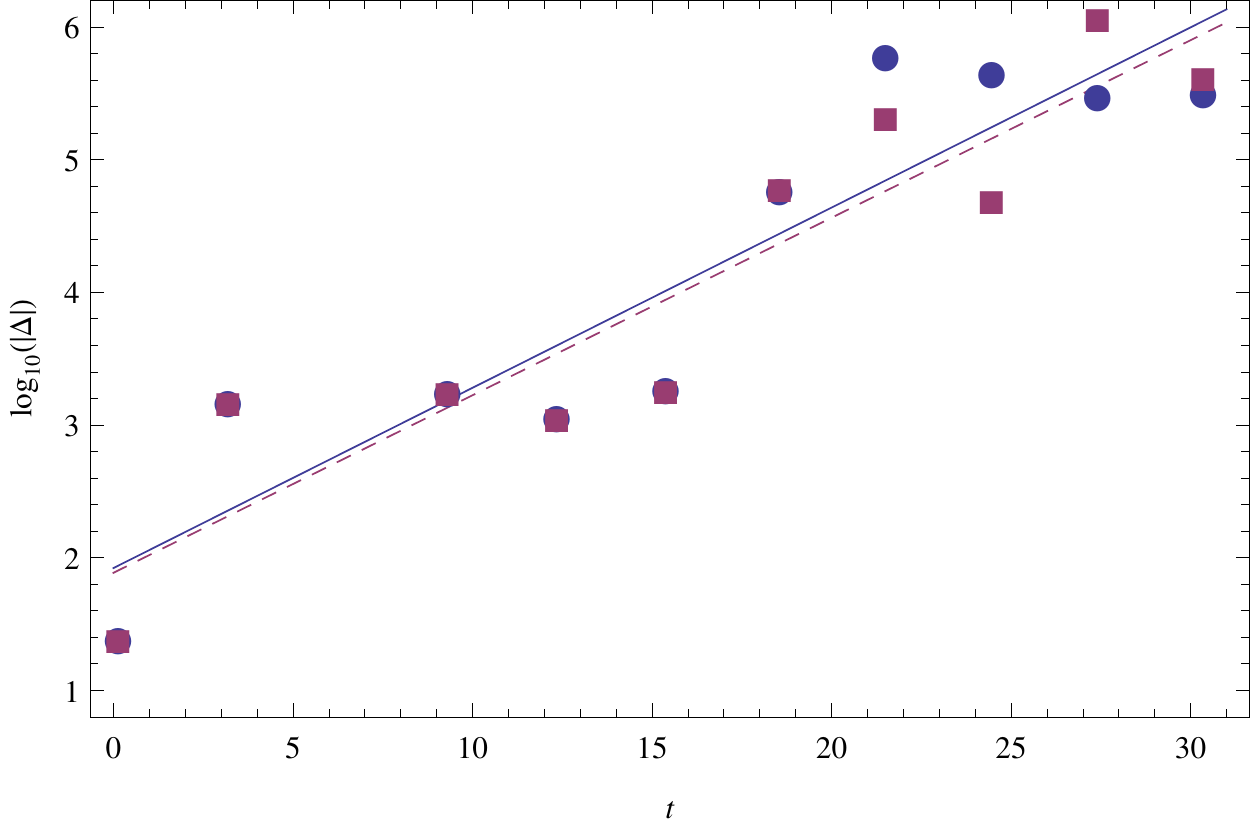}
\caption{$\log|\Delta|$ vs $t$}
\label{fig:lyapExp}
\end{subfigure}
\caption{Left: Upper envelope of Ricci scalar for amplitudes 
$\epsilon_1$ (green dashed), $\epsilon_2$ (blue solid), $\epsilon_3$ 
(red dotted).  Right: Log of the absolute differences $\Delta_{12}$ (blue
points and solid line), $\Delta_{23}$ (magenta squares and dashed line)
and best fits to exponential growth.}
\label{fig:lyapunov}
\end{figure}

Another characteristic of chaotic behavior is exponential growth of 
some measure of distance between two systems with similar initial conditions,
$|\Delta|\sim \exp(\lambda t)$ for Lyapunov exponent $\lambda$.
We consider three neighboring amplitudes in figure \ref{fig:chaosZoom3},
$\epsilon_{1,2,3}=44.413,44.412,44.411$ and take as a measure of the distance
between them the difference in the upper envelope of the Ricci scalar at
the origin $\Delta_{12}=\bar\R_1(t)-\bar\R_2(t)$, etc.  The bar
indicates the maximum (at the origin) over one full reflection from the 
conformal boundary $\Delta t=\pi$.  Figure \ref{fig:lyapRicci} shows the
$\bar\R$ for the three amplitudes (in green dashed, blue solid, and red dotted
curves); note that the three amplitudes lead to different values of $t_H$, so
the $\bar\R_{1,2}$ curves do not extend across the entire plot.
Figure \ref{fig:lyapExp} shows $\Delta_{12}$ (blue points)
and $\Delta_{23}$ (magenta squares) versus time along with best fit exponential
functions (blue solid and magenta dashed lines respectively).  Both fits
are consistent with Lyapunov exponents $\lambda\sim 0.31$, or mild chaotic
behavior.  The actual differences appear to have oscillation superimposed on
the exponential growth.

While \cite{Deppe:2014oua} first suggested that AdS gravitational collapse in
EGB gravity exhibits chaotic behavior, \cite{Brito:2016xvw} have also found
evidence for chaos in the gravitational collapse of two thin shells of 
matter in AdS with Einstein gravity.   As in our figure \ref{fig:chaosZoom1},
\cite{Brito:2016xvw} finds hints for self-similarity in the $t_H$ curve
as a function of initial conditions (in their case, the common initial radius 
of the two shells of matter).  In this system, energy transfers gravitationally
between the shells as they pass through each other; in the self-similar region,
the transfer back and forth leads to chaotic behavior in the horizon formation
time.  The two shells are not both near the origin when the horizon forms; 
instead, the horizon forms when one of the shells happens to have accumulated
a large enough density to form a horizon on its own.
In addition, \cite{Brito:2016xvw} also finds 
a small but positive Lyapunov exponent for the deviations between nearby
initial conditions in the chaotic region of parameter space.  Clearly this
is similar behavior, and there may be a deeper analogy between scalar collapse
in EGB gravity and the two-shell system. Specifically, at least for some
amplitudes, the GB term causes the initial scalar field pulse to break into
multiple pulses, each of which behaves as an independent shell of matter.
For shells with large radii, the GB term is negligible, so we are in fact
also studying the collapse of multiple transparent shells in Einstein gravity.
Examining one of our evolutions as an animation is instructive; 
an animation of $\mass'$ for $\epsilon=44.412$ is available at
\url{http://ion.uwinnipeg.ca/~afrey/AdSGB16.html}.  
We see that the initial pulse slowly separates into two (groups of) pulses
of matter, which are approximately completely out of phase by $t\sim 15$
and each of which contains one tall, thin shell of matter.
Eventually, one of the pulses forms a horizon while the other is far away.
So, once the GB term separates the initial matter distribution into two
pulses, it seems that energy transfer between pulses may be responsible for
the chaotic behavior, as in the two-shell system.  We have also examined
our simulations in the piecewise-constant regions of figure \ref{fig:GBscan}
for comparison;
while collapses that reflect from the boundary multiple times do exhibit
some pulse fragmentation, only the main pulse ever has a high, thin shell of
matter.

\subsection{Naked Singularity Formation}\label{s:naked}

In EGB gravity in AdS$_5$, horizons must contain at least a minimal mass
(even at zero radius); since the asymptotic value of the mass function is
conserved, this implies that horizons cannot form below a critical value of
the amplitude.  For our initial data and choice of GB parameter, 
$\epsilon_{crit}\sim 21.86$.  We have already noted that we have failed to
find horizon formation for $t<100$ for any amplitude $\epsilon\leq 32$,
leaving several important questions.  One, is there a dynamical mechanism
preventing gravitational collapse for small amplitudes that are nontheless
greater than $\epsilon_{crit}$?  We can attempt to answer this by studying
these amplitudes with high-resolution simulations for long times.
For another, do amplitudes that do not form horizons lead to a stable,
quasiperiodic evolution, or can they form naked singularities?  Is the 
behavior the same or different for amplitudes above and below $\epsilon_{crit}$?
To address these questions, we have carried out simulations at 
$\epsilon=20$ and $\epsilon=22$, increasing resolution as necessary to 
carry the simulations to as long a time as possible.  

\begin{figure}[t]
\begin{subfigure}{0.48\textwidth}
\includegraphics[width=0.95\textwidth]{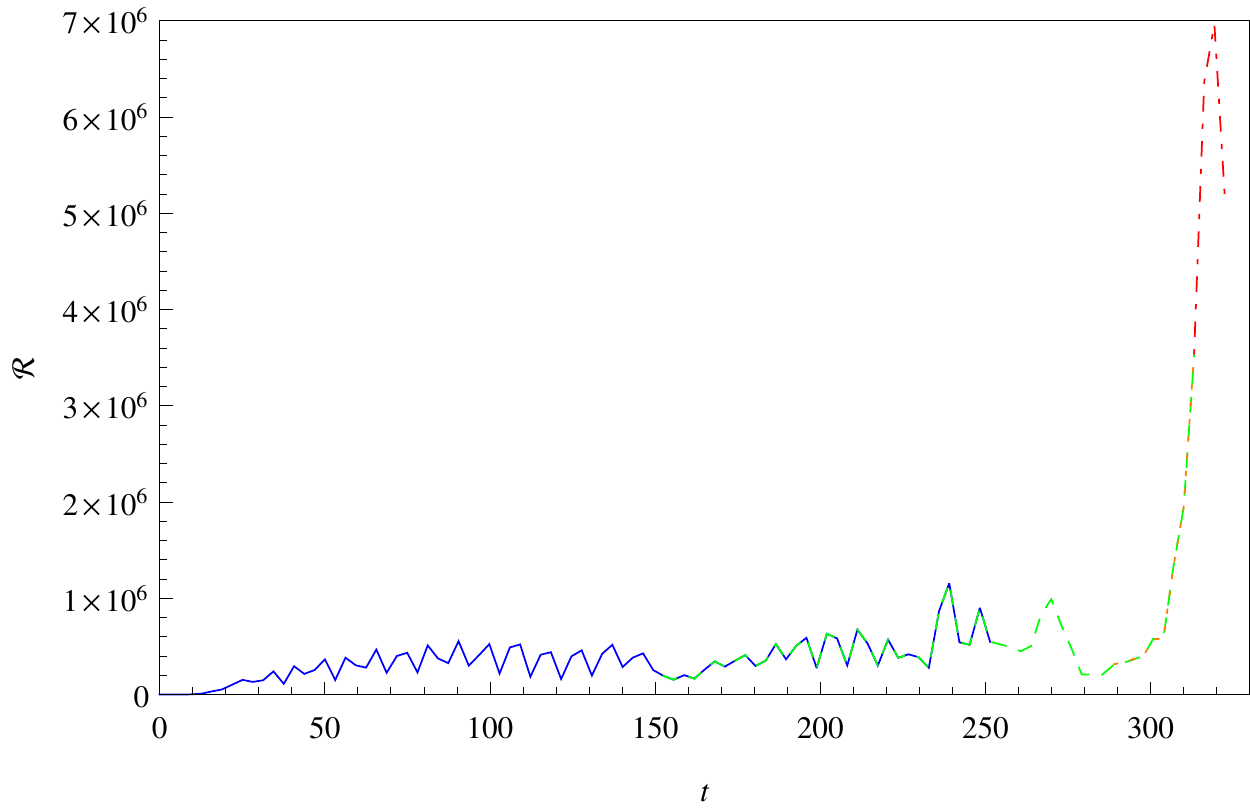}
\caption{Ricci scalar at origin}
\label{fig:ricciA20all}
\end{subfigure}
\begin{subfigure}{0.48\textwidth}
\includegraphics[width=0.95\textwidth]{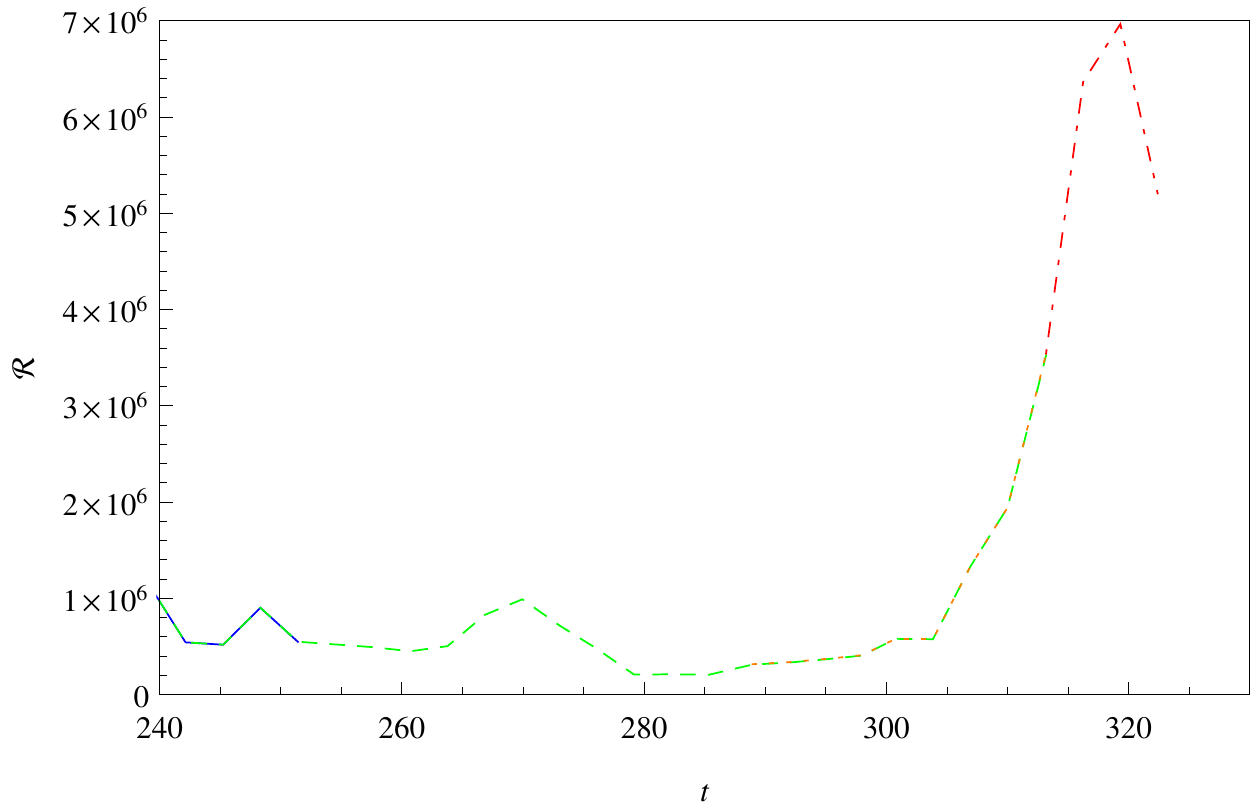}
\caption{Detail of late time behavior}
\label{fig:ricciA20late}
\end{subfigure}
\caption{The upper envelope of Ricci scalar at the origin for amplitude
$\epsilon=20$.  Different curves represent calculations at different
resolutions: $n=15$ (solid blue), $n=16$ (dashed green), $n=17$ (dotted orange),
$n=18$ (dot-dashed red).  The right panel shows
detail for later times in the evolution.}
\label{fig:ricciA20}
\end{figure}

\begin{figure}[t]
\begin{subfigure}{0.48\textwidth}
\includegraphics[width=0.95\textwidth]{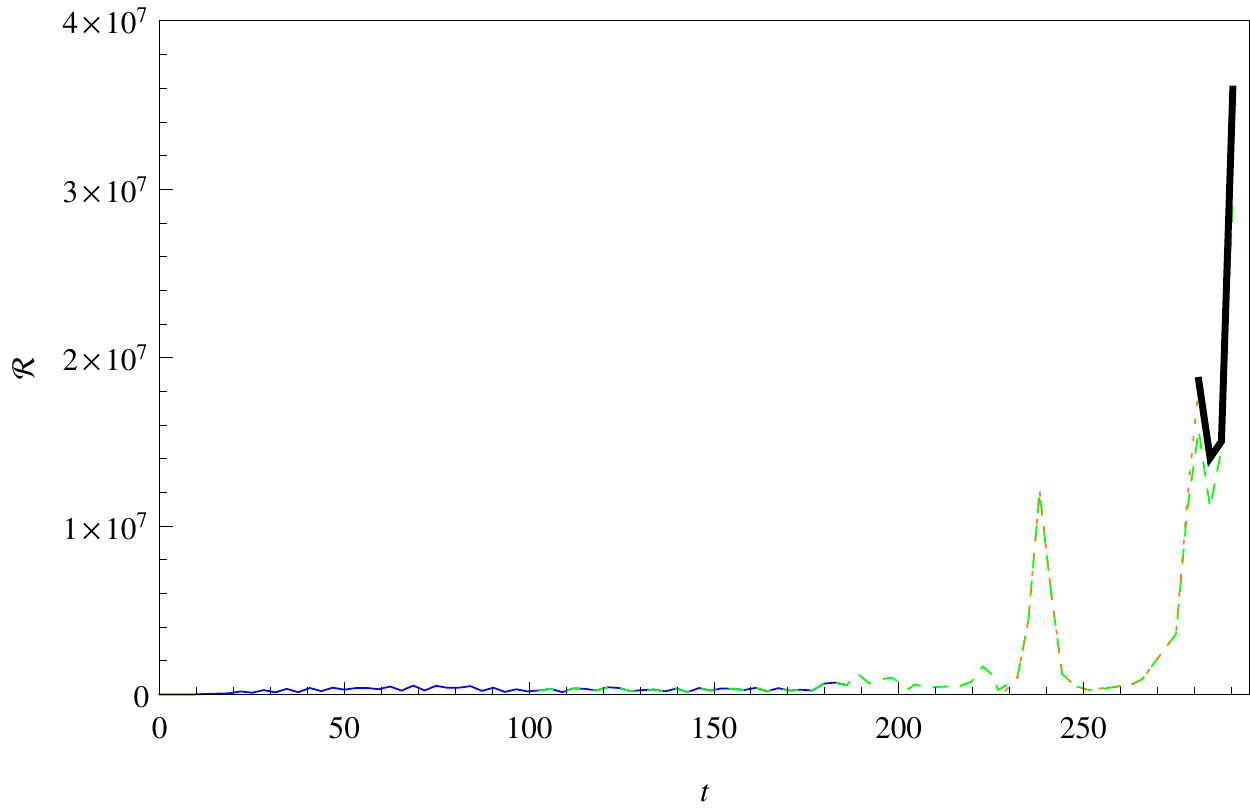}
\caption{Ricci scalar at origin}
\label{fig:ricciA22all}
\end{subfigure}
\begin{subfigure}{0.48\textwidth}
\includegraphics[width=0.95\textwidth]{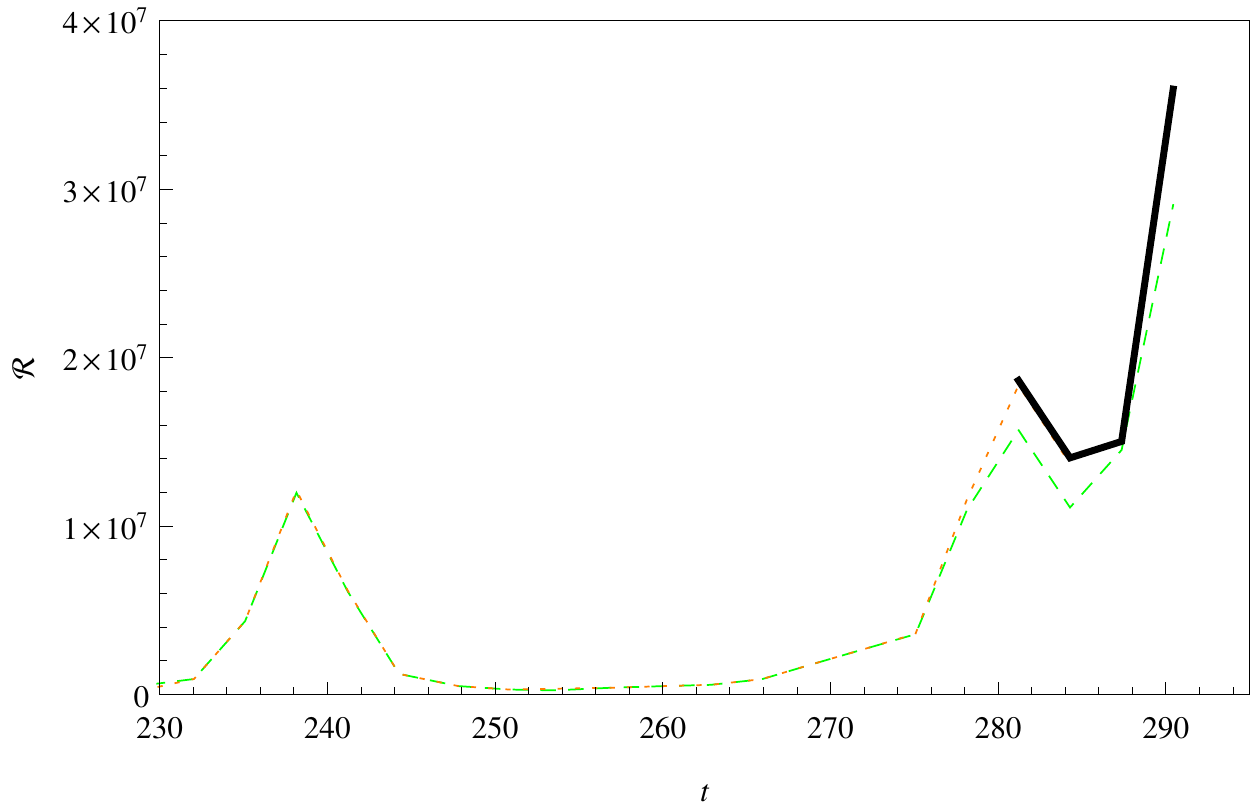}
\caption{Detail of late time behavior}
\label{fig:ricciA22late}
\end{subfigure}
\caption{The upper envelope of Ricci scalar at the origin for amplitude
$\epsilon=22$.  Different curves represent calculations at different
resolutions: $n=15$ (solid blue), $n=16$ (dashed green), $n=17$ (dotted orange),
$n=18$ (dot-dashed red), $n=19$ (thick black).  The right panel shows
detail for later times in the evolution.}
\label{fig:ricciA22}
\end{figure}

We have been unable to find horizon formation in either case to times of 
$t\sim 325,295$ and resolutions up to $n=18,19$ respectively for 
$\epsilon=20,22$. The need for the high resolutions
is clear when we consider $\bar\R$, the upper envelope of the Ricci scalar
at the origin, which we show in figures \ref{fig:ricciA20},\ref{fig:ricciA22}.
In both cases, we find strong growth of the Ricci scalar to very large 
values, eventually reaching values of order $\bar\R\sim 10^7$ while avoiding
formation of a horizon.  From visual inspection of the simulations, the key
physics seems to be dispersal of the original matter pulse into two pulses,
which individually narrow, leading to very high curvatures, but which are
nonetheless not massive enough to form a horizon.  Nonetheless, the extreme
growth of $\bar\R$ and pulse narrowing (which also drives the need for 
increasingly higher resolution at late times) suggests the possibility that
these amplitudes will eventually form a naked singularity, rather than 
behaving in a quasi-periodic fashion.

\begin{figure}[t]
\begin{subfigure}{0.48\textwidth}
\includegraphics[width=0.95\textwidth]{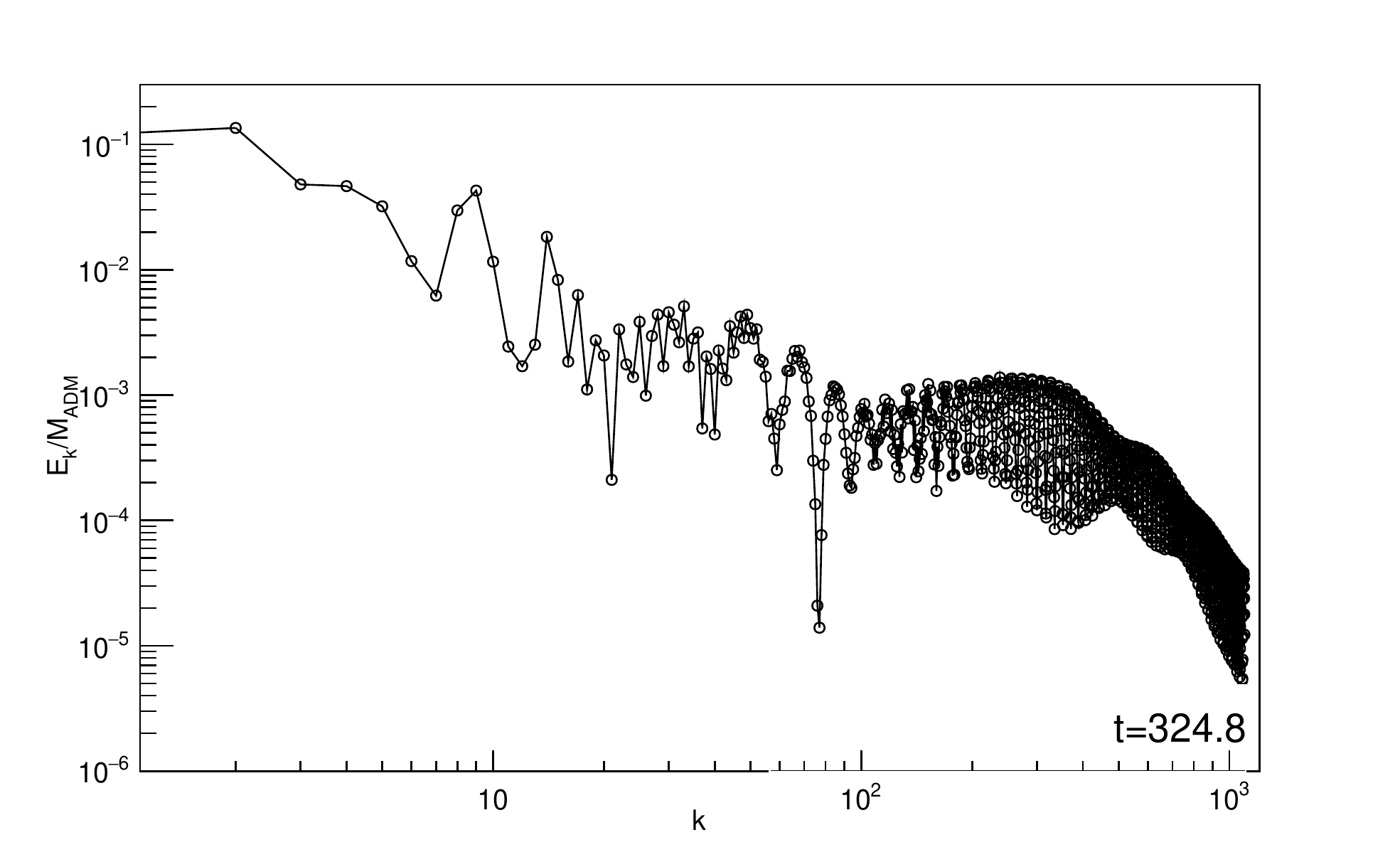}
\caption{$\epsilon=20$}
\label{fig:spectrumA20}
\end{subfigure}
\begin{subfigure}{0.48\textwidth}
\includegraphics[width=0.95\textwidth]{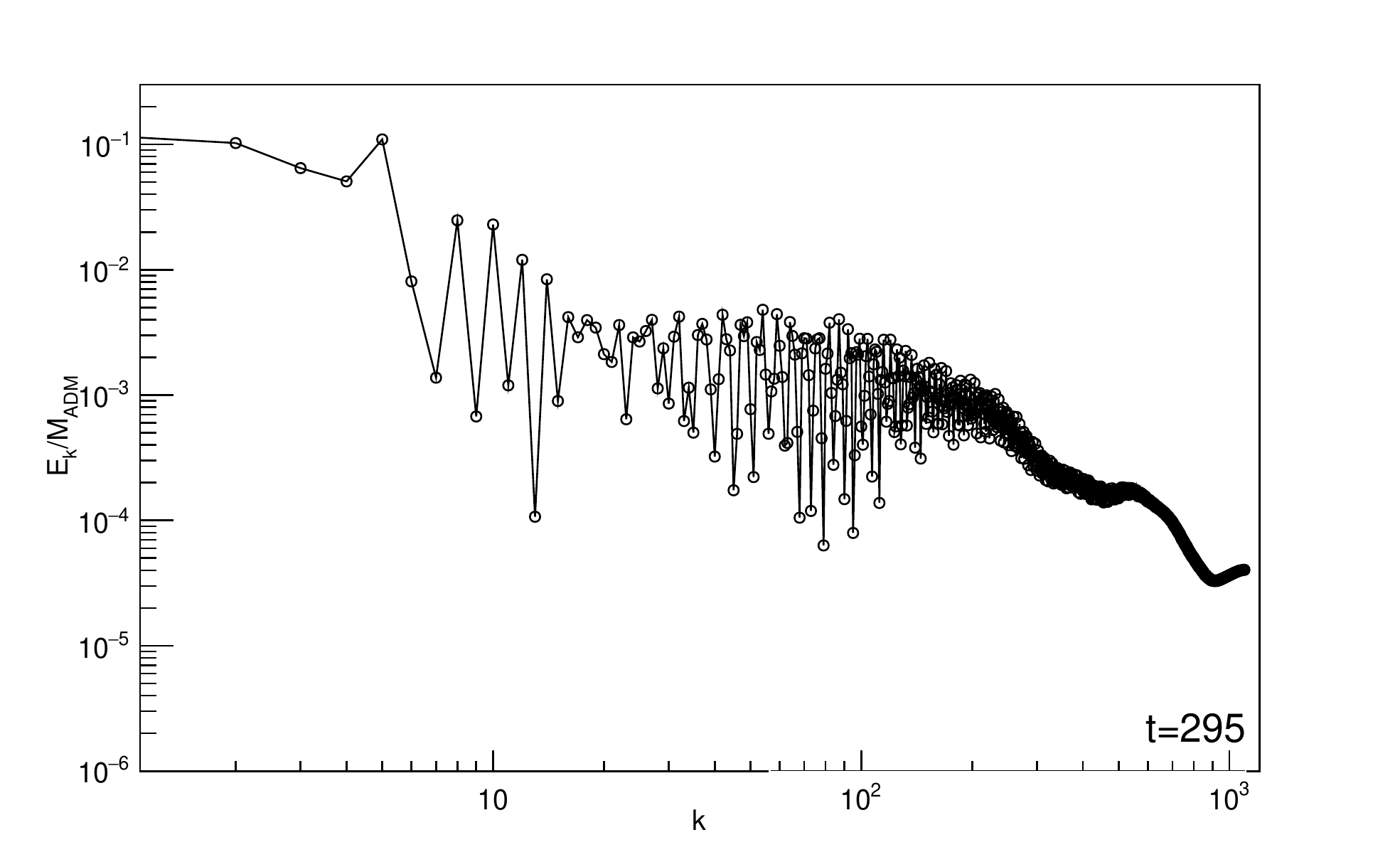}
\caption{$\epsilon=22$}
\label{fig:spectrum22}
\end{subfigure}
\caption{Energy spectra as fraction of total energy per mode at the latest
times simulated.  Spectra are derived from simulations with $n=18$ 
resolution in both cases for computational resource reasons.}
\label{fig:spectra}
\end{figure}

As further suggestive evidence of singularity formation, we have studied the
late-time energy spectra of both evolutions.  Figure \ref{fig:spectra} shows
the energy spectra (to the $j=1024$ eigenmode) 
for both amplitudes at the latest time we were able to simulate in each case. 
These show a slow power-law decay at large mode number, which is usually
characteristic of horizon formation.\footnote{To our knowledge the first
demonstration of a power law spectrum in gravitational collapse in AdS
was given in \cite{deOliveira:2012dt} for the Fourier modes of the Ricci
scalar at the origin near horizon formation.}  
However, it is impossible for a 
horizon to form for $\epsilon=20$, so the distribution of energy through
the higher modes suggests the possible development of a naked singularity.
Another point suggestive of singularity formation is that we find over 1\% of 
the total energy lies in higher modes ($j>1024$) for times greater than
$t\sim 322.4$.  Similarly, the $\epsilon=22$ evolution seems to be moving
rapidly toward either horizon or naked singularity formation for $t\sim 295$.
Over 1\% of the total energy is in higher modes for $t\gtrsim 287$,
and close to 3\% is in higher modes by the end of our simulation at 
$t\sim 295$.  This degree of energy in high eigenmodes allows an extreme
concentration of energy density near the origin, which could drive 
the formation of a singularity.  It is important to note that these spectra
differ from that at earlier times (see for example the supplemental
information of \cite{Deppe:2014oua}), which have an apparent exponential 
cut-off for mode numbers less than $10^3$, indicating that the power has 
continued to cascade into higher and higher modes as the evolution progressed.
This difference is one reason for a potentially different conclusion about
evolution at amplitudes near $\epsilon_{crit}$ in comparison to 
\cite{Deppe:2014oua}.

Assuming that a naked singularity does form at a finite time $t$, which is
the proper time at the origin, it is important to ask whether redshift
effects push the singularity formation to an infinite conformal time at the
boundary, which controls the physics in the dual CFT.  Unlike the case
of horizon formation, however, time dilation effects seem to be unimportant
here; at the latest times probed by our simulations, the metric function
$\delta$ takes values of $-0.007$ and $-0.035$ at the boundary 
for $\epsilon=20,22$ respectively, which are the minima over their evolution.

It is worth considering at this point what the formation of a naked singularity
at finite time would mean in the context of the AdS/CFT correspondence.
If the gravity side of the correspondence is described by the pure EGB
gravity with no additional higher-curvature terms, the translation of a
naked singularity to the dual field theory is unclear.  It is tempting,
therefore, to postulate that a naked singularity is a sign of a pathology
in the theory; in fact, \cite{Camanho:2014apa} has already argued that
theories dual to pure EGB gravity in AdS suffer from acausalities.  On the
other hand, if the Gauss-Bonnet term is just the first in a tower of 
higher-curvature corrections, the extreme curvatures of figures 
\ref{fig:ricciA20},\ref{fig:ricciA22} suggest that the additional corrections
will become significant (as seems to be the case in the self-similar 
transition regions of section \ref{s:chaotic} as well).  That would be 
a signal that the effective gravity theory is breaking down and should be
replaced by the full string theory, and it is possible that the end state of
collapse is a gas of strings near the origin.\footnote{We thank A.~Buchel
for interesting discussions related to this point.}

Due to the very high resolutions necessary, it was not computationally 
feasible to perform convergence testing for the entire simulations shown in
figures \ref{fig:ricciA20},\ref{fig:ricciA22}.  However, convergence tests
for part of the $\epsilon=20,n=17$ calculations showed the expected 4th-order
convergence for the Ricci scalar at the origin, and the clear overlap of
the different resolutions in much of the figures argues for the reliability of
our results.

\section{Discussion}\label{s:conclusions}

In this manuscript, we have expanded on the analysis of black hole formation
in AdS Einstein-Gauss-Bonnet gravity first presented in 
\cite{Deppe:2014oua}.  In our examination of the horizon formation time $t_H$
as a function of amplitude (figure \ref{fig:GBscan}), we have considered
three particular physical phenomena in detail: critical behavior at
a transition in $t_H$, possibly chaotic behavior below the transition
amplitude, and long-time evolution at low amplitudes possibly hinting at
formation of naked singularities.

We first examined critical behavior at the transition from immediate collapse,
i.e. when a black hole forms without the matter first dispersing, to collapse
after one or more reflections from the conformal boundary
and confirmed the existence of a
dynamical radius gap of approximately $R_{min}\sim 10^{-1.9}$ (see
figure \ref{fig:GBchoptuik}).  A number of questions remain about this
critical behavior.  For example, do the transitions from other
step regions of figure \ref{fig:GBscan} exhibit the same radius gap as in
figure \ref{fig:GBchoptuik}, or does the value change (or even vanish)?
Is there a universal scaling law for the horizon radius for amplitudes
\comment{below the jumps in \ref{fig:StepsInScaling} or}
below any transitions in 
figure \ref{fig:GBscan}, as was demonstrated in Einstein gravity in
\cite{Olivan:2015fmy,Santos-Olivan:2016djn}? We expect that the
radius gap is due to the existence of a massive critical 
solution, in which case the local features of the critical behaviour
should be independent of the number of reflections from the boundary before
horizon formation. Thus, the radius gap should persist for higher numbers of
bounces. Some evidence in this direction was provided in figure 3 of
\cite{Deppe:2014oua}, which compares the scaling plots after one bounce
to those with no bounces.

Our simulations also shed light on a novel dynamical feature
of the critical behavior, first presented in \cite{Deppe:2014oua}. The
step-like behaviour in the scaling plot (figure \ref{fig:StepsInScaling})
a time dilation effect; part of the initial lump of matter disperses from
the origin rather than falling into the forming horizon.  For amplitudes
close enough to the transition value, one or more of these sub-pulses have
enough time to reflect from the AdS boundary and return to the origin before
the simulation reaches our criterion for horizon formation.  Although not
previously observed, this could in principle occur in any gravity theory
in AdS that exhibits critical behaviour, including Einstein
gravity.  It would be interesting to check since this effect depends both
on global features of AdS and local dynamics.

The transitions from one piecewise-constant region of figure \ref{fig:GBscan}
to another also demonstrate significant scatter.  We have presented
evidence that the $t_H$ vs $\epsilon$ curve is self-similar in the
region from $\epsilon\sim 44.0$ to 45.3.  Furthermore, a positive Lyapunov
exponent between the evolutions of nearby amplitudes is a hallmark of
truly chaotic behavior.  The chaotic behaviour appears to have as its
source the separation of the initial pulse into two (or possibly more)
pulses and the subsequent transfer of energy between them as they pass
through each other between reflections off infinity and the
origin. Here, too, questions remain: Is chaos present for any
gravitational system in which the matter forms multiple pulses, as
is the case in the two-shell system of \cite{Brito:2016xvw}? Does the 
matter distribution fragment whenever the physics has a second scale
(other than the AdS scale), such as a mass for the scalar field?

Perhaps the most intriguing aspect of our analysis is the potential
evidence for naked singularity formation in the model. Below the
algebraic mass gap of $M_{crit}=\lambda_3/2$ ($\epsilon_{crit} = 21.86$ 
for our GB parameter and initial data), 
there are only two possible end states: a
quasi-periodic steady state or naked singularity formation. While it
is impossible to provide definitive proof numerically, the observed
dramatic increase in curvature and concentration of energy into higher
modes in the absence of horizon formation suggest that the
end state will be a naked singularity for our two long evolutions near the
critical value (one just below, the other just above).  In pure EGB 
gravity, naked singularity formation may indicate a pathology of any dual
field theory.  On the other hand, the extreme curvatures found may instead
indicate the excitation of other higher curvature terms and 
string degrees of freedom, leading to the eventual production
of a gas of strings rather than the actual development of a singularity.

\begin{figure}[t]
\begin{subfigure}{0.48\textwidth}\ \\ \ \\
\includegraphics[width=0.88\textwidth]{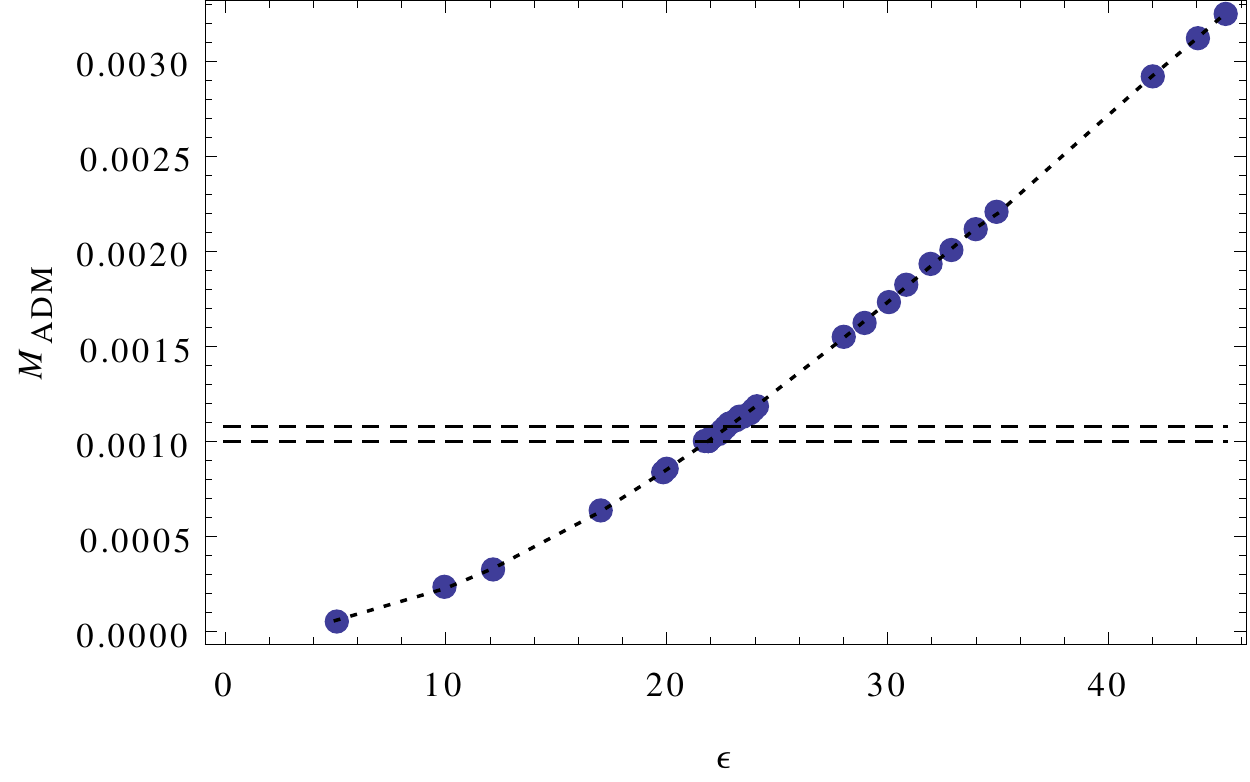}
\caption{ADM Mass vs amplitude}
\label{fig:massAmp}
\end{subfigure}
\begin{subfigure}{0.48\textwidth}
\includegraphics[width=0.98\textwidth]{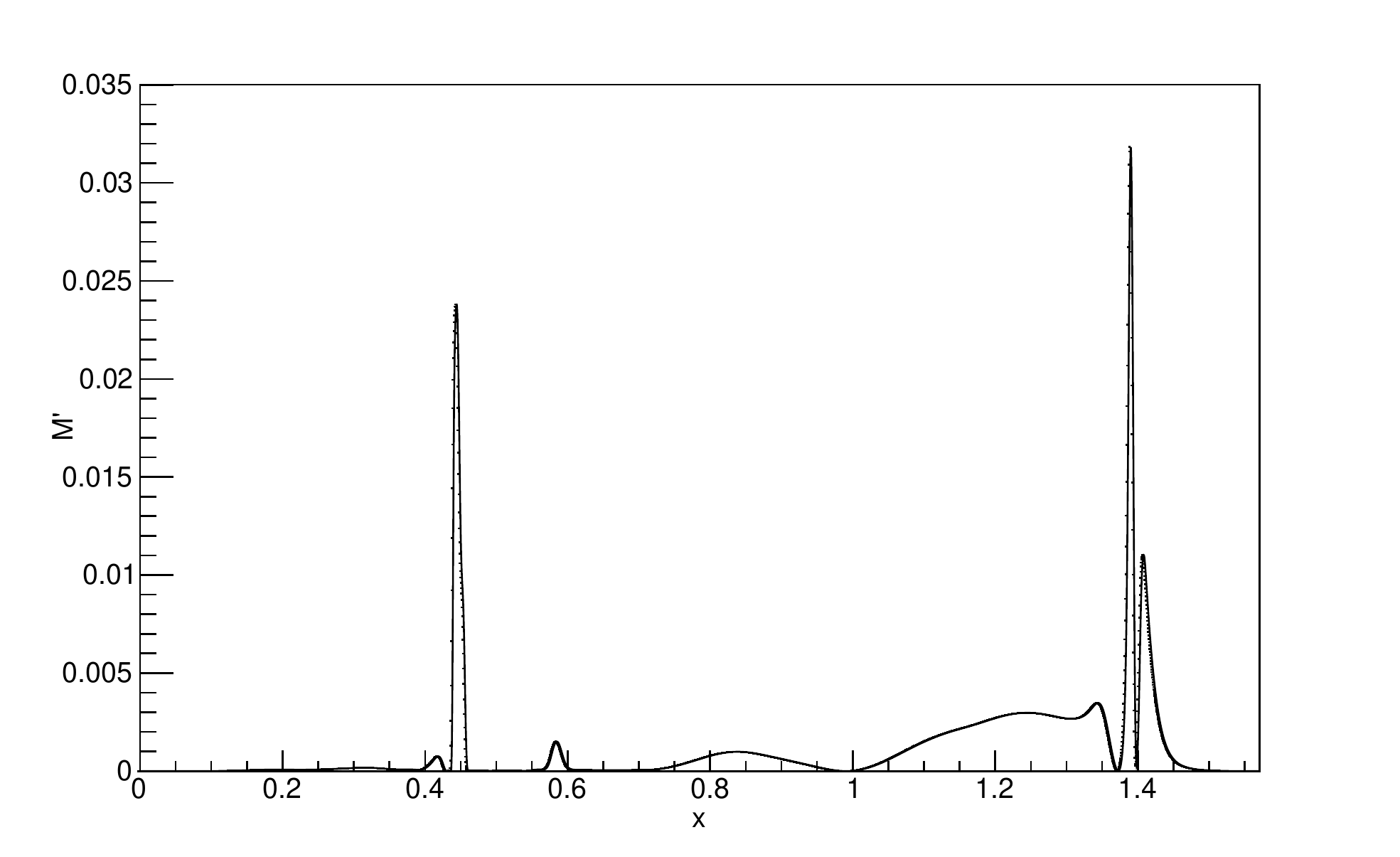}
\caption{Mass in two pulses at low amplitude}
\label{fig:twopulses}
\end{subfigure}
\caption{Left: The conserved mass as a function of the amplitude
of initial data, including gravitational effects, as calculated from 
simulations.  The dashed horizontal lines indicate $M_{crit}$ (lower line)
and the dynamical limiting mass $\mass(R_{min})$ (upper line). Right:
$M'$ vs $x$ at $t=93.99$ for the evolution of $\epsilon=30.2$ initial data.
There are two separated thin shells of matter.}
\label{fig:suppression}
\end{figure}

It is curious that horizon formation seems strongly
suppressed (takes a lot longer, or does not form at all) even above
the algebraic threshold $\epsilon_{crit}$.  Our results allow us to speculate
as to the cause of this suppression, which seems to be a highly nonlinear
effect.  In the case of our evolution just above threshold,
$\epsilon = 22$, the ADM mass is $M\sim 0.00101$, which is just barely
above the critical value (see figure \ref{fig:massAmp} for the conserved
mass as a function of amplitude). 
Assuming the existence of a dynamical radius
gap of about $R_{min}\sim 10^{-1.9}$, equation (\ref{eq:massathorizon}) 
implies that the minimum amount of
mass dynamically required to form a horizon is actually close to
$\mass(R_{min})\sim 0.00108$, so both evolutions discussed in 
section \ref{s:naked} were below this dynamical limit. The
fact that we do not see a black hole form slightly above the critical
value is perhaps not a surprise.  More surprising is the
apparent suppression of black hole formation for amplitudes near
$\epsilon = 32$ and below (see figure \ref{fig:GBscan}). 
The mass at this amplitude
is close to 0.002, double $M_{crit}$.  However, we have seen in chaotic
transition regions that the mass tends to split into at least
two pulses. In this case, the splitting can
produce smaller shells of matter that individually do not have enough energy to
form a horizon, so that horizon formation depends on the subsequent
energy transfer between pulses/shells. Indeed, this splitting occurs at 
long times for $\epsilon=30.2$, as indicated in figure \ref{fig:twopulses},
so it seems that the lowest amplitudes shown in figure \ref{fig:GBscan}
are in a chaotic region.  Since the two pulses appear to carry a substantial
fraction of the mass, horizon formation will require significant energy
transfer between pulses (more than in chaotic regions at higher amplitude).
It seems likely that this significant transfer is unlikely and will occur
only rarely, leading to very long horizon formation times.  It is also
worth speculating if similar physics occurs for gravitational collapse in 
AdS$_3$, which also has a critical black hole mass and apparent suppression
of horizon formation for amplitudes just above the critical mass
\cite{Bizon:2013xha,daSilva:2014zva}.

It is clear that gravitational collapse in EGB gravity in AdS is an 
intricate system, showing first-order transitions, chaotic behavior, and
possible formation of naked singularities.

\acknowledgments

We gratefully acknowledge many helpful conversations with P.~Bizo\'n,
A.~Buchel, and S.~Green.
The work of ND is supported in part by a Natural Sciences and Engineering
Research Council of Canada PGS-D grant to ND, NSF Grants PHY-1306125
and AST-1333129 at Cornell University, and by a grant from the Sherman
Fairchild Foundation.
The work of AF and GK is supported by the Natural Sciences
and Engineering Research Council of Canada Discovery Grant program.  
This research was enabled in part by support provided by WestGrid 
(www.westgrid.ca) and Compute Canada Calcul Canada (www.computecanada.ca).

\appendix
\section{Appendix: Generalized Flat Slice (PG) Coordinates} 
%\subsection{Choice of Metric Function}
Working as before with $R = R(x)$, 
we choose:
\be
\Lambda = \Lambda(x)
\ee
Flat slice coordinates would correspond to the choice  $\Lambda =1$, 
but this is not appropriate when the cosmological constant is non-zero 
\cite{Soo2009}. Instead we first write the mass function as
\bea
\mass &=& \frac{n-2}{2\kappa_n^2}R^{n-1}\left[\lambda + \frac{1}{R^2} 
\left(1-\left(\frac{R^\prime}{\Lambda}\right)^2\right)
+\frac{\lambda_3}{R^4}\left(1-\left(\frac{R^\prime}{\Lambda}\right)^2\right)^2
\right.\nonumber\\
 & & \left.   + \left[ \frac{1}{R^2}+2\frac{\lambda_3}{R^4}\left(1-\left(
\frac{R^\prime}{\Lambda}\right)^2\right)\right]y^2+\frac{\lambda_3}{R^4}y^4 
\right]
\eea
and choose a function $\Lambda(x)$ that yields a diagonal  metric in vacuum.
This requires the first three terms in the square brackets above to vanish, 
which in turn implies that
\be
\Lambda(x) = \frac{R^\prime}{\sqrt{1+\lambda_{eff} R^2}}\ .
\label{eq:LambdaPG}
\ee
The sign in front of the square root is chosen to give the usual answer 
$\Lambda = R^\prime$ when $\lambda=0$. 
When $\lambda_3=0$ and $R=x$, this gives:
\be
\Lambda = \left(1+ {\lambda}R^2\right)^{-1/2} \ ,
\ee
which is correct for AdS in Schwarzschild coordinates for $M=0$.

With this choice the mass function reduces to:
\bea
M &=& \frac{3}{2\kappa_5^2}R^{2}\left[ \left[ 1+2\frac{\lambda_3}{ R^2}
\left(1-\left(\frac{R^\prime}{\Lambda}\right)^2\right)\right]y^2+
\frac{\lambda_3}{ R^2}y^4 \right]\nonumber\\
&=& \frac{3}{2\kappa_5^2}R^{2}\left[ \left[ 1-2{\lambda_3}\tilde{D} )\right]y^2
+\frac{\lambda_3}{ R^2}y^4 \right]\ .
\label{eq:MPG2}
\eea
The Hamiltonian constraint is
\begin{equation}
\label{eq:2}
M^\prime =\frac{R^\prime}{\Lambda^2}\rho_m-\frac{y}{\Lambda}P_\psi\psi^\prime
\ ,\end{equation}
which determines $y= - (N_r/N) R'$ in terms of data on a slice.
In these coordinates the consistency condition $\dot{\Lambda}=0$ is found to be
\begin{equation}
\label{eq:consistencyPG}
\left(\frac{N\Lambda}{R^\prime}\right)^\prime=
\frac{N}{R^\prime}P_\psi\psi^\prime
\left(\left.\frac{\partial M}{\partial y}\right|_\Lambda\right)^{-1}\ ,
\end{equation}
where using (\ref{eq:MPG2})
\bea
\left(\left.\frac{\partial M}{\partial y}\right|_\Lambda\right)
&=& \frac{3}{2\kappa_5^2}R^{2}\left[2 \left[ 1-2{\lambda_3}\lambda_{eff})\right]
y+4\frac{\lambda_3}{ R^2}y^3\right]
\label{eq:dMdy}
\eea
The dynamical equations in terms of $\Pi = P_\psi/R^{n-2}$ and
$\Phi= \psi^\prime$ are
\bea
  \dot{\Phi}
  &=&\left(N\left[\frac{\Pi}{\Lambda }
  -\frac{y}{R^\prime}
\Phi\right] \right)^\prime
    \label{eq:Phi dot}\\
  \dot{\Pi}   &=&\frac{1}{R^{3}}\left[NR^{3}\left(\frac{\Phi}{\Lambda}
-\frac{y\Pi}{R^\prime}
\right)\right]^\prime \ .
 \label{eq:Pi dot}
 \eea

\bibliographystyle{JHEP}
\bibliography{library}
\end{document}